\documentclass[twocolumn,mathlines,collaboration]{aastex631}
\usepackage[utf8]{inputenc}
\usepackage{graphicx}
\usepackage{color}
\usepackage{amsmath}
\usepackage{comment}
\usepackage[first=0,last=9]{lcg}

\usepackage{array,multirow}

\hypersetup{linkcolor=blue,citecolor=blue,filecolor=blue,urlcolor=blue}

\usepackage[acronym]{glossaries}
\setacronymstyle{long-short}
\glsdisablehyper
\newacronym{llama}{LLAMA}{Low-Latency Algorithm for Multimessenger Astrophysics}

\begin{document}

\title{Deep Search for Joint Sources of Gravitational Waves and High-Energy Neutrinos with IceCube During the Third Observing Run of LIGO and Virgo}
\author[0000-0001-6141-4205]{R. Abbasi}
\affiliation{Department of Physics, Loyola University Chicago, Chicago, IL 60660, USA}

\author[0000-0001-8952-588X]{M. Ackermann}
\affiliation{Deutsches Elektronen-Synchrotron DESY, Platanenallee 6, D-15738 Zeuthen, Germany}

\author{J. Adams}
\affiliation{Dept. of Physics and Astronomy, University of Canterbury, Private Bag 4800, Christchurch, New Zealand}

\author[0000-0002-9714-8866]{S. K. Agarwalla}
\altaffiliation{also at Institute of Physics, Sachivalaya Marg, Sainik School Post, Bhubaneswar 751005, India}
\affiliation{Dept. of Physics and Wisconsin IceCube Particle Astrophysics Center, University of Wisconsin{\textemdash}Madison, Madison, WI 53706, USA}

\author[0000-0003-2252-9514]{J. A. Aguilar}
\affiliation{Universit{\'e} Libre de Bruxelles, Science Faculty CP230, B-1050 Brussels, Belgium}

\author[0000-0003-0709-5631]{M. Ahlers}
\affiliation{Niels Bohr Institute, University of Copenhagen, DK-2100 Copenhagen, Denmark}

\author[0000-0002-9534-9189]{J.M. Alameddine}
\affiliation{Dept. of Physics, TU Dortmund University, D-44221 Dortmund, Germany}

\author[0009-0001-2444-4162]{S. Ali}
\affiliation{Dept. of Physics and Astronomy, University of Kansas, Lawrence, KS 66045, USA}

\author{N. M. Amin}
\affiliation{Bartol Research Institute and Dept. of Physics and Astronomy, University of Delaware, Newark, DE 19716, USA}

\author[0000-0001-9394-0007]{K. Andeen}
\affiliation{Department of Physics, Marquette University, Milwaukee, WI 53201, USA}

% ...

\collaboration{434}{The IceCube Collaboration}
\author[0000-0003-4786-2698]{A.~G.~Abac}
\affiliation{Max Planck Institute for Gravitational Physics (Albert Einstein Institute), D-14476 Potsdam, Germany}

\author{R.~Abbott}
\affiliation{LIGO Laboratory, California Institute of Technology, Pasadena, CA 91125, USA}

\author{I.~Abouelfettouh}
\affiliation{LIGO Hanford Observatory, Richland, WA 99352, USA}

\author{F.~Acernese}
\affiliation{Dipartimento di Farmacia, Universit\`a di Salerno, I-84084 Fisciano, Salerno, Italy}
\affiliation{INFN, Sezione di Napoli, I-80126 Napoli, Italy}

\author[0000-0002-8648-0767]{K.~Ackley}
\affiliation{University of Warwick, Coventry CV4 7AL, United Kingdom}

\author{S.~Adhicary}
\affiliation{The Pennsylvania State University, University Park, PA 16802, USA}

\author[0000-0002-4559-8427]{N.~Adhikari}
\affiliation{University of Wisconsin-Milwaukee, Milwaukee, WI 53201, USA}

\author[0000-0002-5731-5076]{R.~X.~Adhikari}
\affiliation{LIGO Laboratory, California Institute of Technology, Pasadena, CA 91125, USA}

\author{V.~K.~Adkins}
\affiliation{Louisiana State University, Baton Rouge, LA 70803, USA}

\author[0000-0002-8735-5554]{D.~Agarwal}
\affiliation{Universit\'e catholique de Louvain, B-1348 Louvain-la-Neuve, Belgium}
\affiliation{Inter-University Centre for Astronomy and Astrophysics, Pune 411007, India}

% ...

\collaboration{1771}{The LIGO Scientific Collaboration, The Virgo Collaboration, and The KAGRA Collaboration}

\begin{abstract}

The discovery of joint sources of high-energy neutrinos and gravitational waves has been a primary target for the LIGO, Virgo, KAGRA, and IceCube observatories. The joint detection of high-energy neutrinos and gravitational waves would provide insight into cosmic processes, from the dynamics of compact object mergers and stellar collapses to the mechanisms driving relativistic outflows. The joint detection of multiple cosmic messengers can also elevate the significance of the common observation even when some or all of the constituent messengers are \emph{sub-threshold}, i.e. not significant enough to declare their detection individually. Using data from the LIGO, Virgo, and IceCube observatories, including sub-threshold events, we searched for common sources of gravitational waves and high-energy neutrinos during the third observing run of Advanced LIGO and Advanced Virgo detectors. Our search did not identify significant joint sources. We derive constraints on the rate densities of joint sources. Our results constrain the isotropic neutrino emission from gravitational-wave sources for very high values of the total energy emitted in neutrinos ($>10^{52}$--$10^{54}$~erg).

\end{abstract}

\section{Introduction} \label{sec:Introduction}

With the discoveries of new cosmic messengers, multimessenger astrophysics has become a reality. Astrophysical high-energy ($\gtrsim$TeV) neutrinos (HENs) were discovered in 2013 \citep{2013Sci...342E...1I,2014PhRvL.113j1101A} by the IceCube Neutrino Observatory~\citep{Aartsen_2017,Aartsen_2024}(IceCube in the following). Multimessenger science matured with the gravitational-wave~(GW) discoveries by Advanced LIGO~\citep{2015CQGra..32g4001L} and Advanced Virgo~\citep{2015CQGra..32b4001A}, including the detection of binary black hole (BBH) coalescences \citep{Abbott_2016,2016PhRvX...6d1015A,2017PhRvL.118v1101A,2017PhRvL.119n1101A} and a binary neutron star coalescence (BNS) in their first two observing runs~\citep{2017PhRvL.119p1101A,2019PhRvX...9a1001A}. IceCube has been taking data in its full 86-string configuration continuously since 2011~\citep{2017JInst..12P3012A} overlapping with the LIGO Scientific, Virgo, and KAGRA (LVK) Collaboration~\citep{2019PhRvX...9c1040A,2021PhRvX..11b1053A,2021arXiv210801045T,2023PhRvX..13d1039A,2025arXiv250818082T} observing runs, which have included the discovery of 218 probable cosmic compact binary sources. The multimessenger observation of the merging BNS system GW170817~\citep{2017PhRvL.119p1101A} established connections between GW sources and emissions in the electromagnetic spectrum from gamma-rays to radio~\citep{2017ApJ...848L..12A}. 

What was missing from the observed multimessenger repertoire of GW170817 were HENs which were expected but not observed~\citep{2017ApJ...850L..35A,2019ARNPS..69..477M,2013RvMP...85.1401A}. 
HENs are predicted to be emitted from the jets forming after the time of GW emission. These jets are expected to accelerate charged particles, leading to the production of mesons, and the subsequent decay of these mesons should result in the emission of detectable HENs~\citep{2013RvMP...85.1401A,2017ApJ...849..153F,2018PhRvD..98d3020K}. Consequently detection of HENs will carry information about the hadronic processes consequent to GW emissions. Short gamma-ray bursts (GRBs) have also been expected from such jets which was observationally confirmed with GW170817. However, despite this expected connection between GRBs and HENs, the searches for HENs coincident with GRBs have not yielded any significant detection. The searches constrained the HEN emission from GRBs, with stricter constraints for long GRBs~\citep{Aartsen_2017_gamma,Abbasi_2022_gamma,Abbasi_2024_gamma,Abbasi_2024_gamma_err}. The searches for HENs coincident with GWs can also illuminate the GRB-HEN relation better, thanks to the observed connection between BNS mergers and short GRBs.

The large overlap between the LVK observation schedules and the nearly-continuous operation of IceCube allows searches for common sources of HENs and GWs. Consequently, multiple searches have been executed, including some before the discovery of astrophysical GWs~\citep{2013JCAP...06..008A, 2014PhRvD..90j2002A,2016PhRvD..93l2010A,2017ApJ...850L..35A,2017PhRvD..96b2005A,2019ICRC...36..930K,2019ICRC...36..865D,2019ApJ...870..134A,2020EPJC...80..487A,2020ApJ...898L..10A,2022icrc.confE.950V,2023ApJ...944...80A,2023ApJ...959...96A}. IceCube~\citep{2021arXiv210513160A}, Super-Kamiokande~\citep{Abe_2016,Abe_2018,2021ApJ...918...78A}, KamLAND~\citep{2021ApJ...909..116A}, and Borexino~\citep{2017ApJ...850...21A} also searched for coincident low-energy neutrinos. While the successive searches have produced better and better constraints on source populations, the improving quality and growing size of the data did not result in a confident multimessenger discovery with neutrinos and GWs together. 

With the growing interest in such searches~\citep{2013RvMP...85.1401A} and their results~\citep{2021ApJ...909..126K}; methodology~\citep{2008CQGra..25k4039A,2009IJMPD..18.1655V,PhysRevLett.107.251101,2012PhRvD..85j3004B,2019PhRvD.100h3017B,2019arXiv190105486C,Veske_2020,2021ApJ...908..216V,2022APS..APRK14005M} and theory have progressed~\citep{2018PhRvD..98d3020K,2019ApJ...878...34F,2020ApJ...905L..13D,2020MNRAS.492..843G}, forming a firm foundation for the comprehensive search for common sources of GWs and high-energy neutrinos we describe in this paper.

Cosmic phenomena accompanied by accretion can cause acceleration of hadrons. HENs can provide information about the acceleration mechanisms and their environments ~\citep{2002RPPh...65.1025H,2003PhRvD..68h3001R,2012PhRvD..86h3007B,2006JCAP...05..003L}. Indeed, IceCube observed neutrinos from a range of origins, including a diffuse flux potentially with unresolved point sources ~\citep{2013PhRvL.111b1103A,2013Sci...342E...1I}, the blazar TXS 0506+056 ~\citep{2018Sci...361.1378I,2018ApJ...864...84K,2018ApJ...863L..30A}, the active galaxy NGC 1068~\citep{ngc1068} and from the Milky Way~\citep{galacticplane}. However, the majority of observed IceCube cosmic neutrinos are not associated with any identified source. This motivates studies which focus on neutrino emission from observable and well-characterizable sources, such as gravitational-waves sources. 

Discovery and observation of joint sources of HENs and GWs will have a multifold impact, including a better understanding of emitter physics, source localization, and messenger characterization ~\citep{2013RvMP...85.1401A,2017ApJ...849..153F,2018PhRvD..98d3020K}. A coincident detection with a GW event can shed light on the jet physics involved. For example, a coincidence without a gamma-ray counterpart in a binary neutron star merger can indicate presence of a choked jet. In addition there are different estimations for the jet energetics. Searches for joint emissions constrain possible jet characteristics with or without detections. HEN observations can also significantly constrain the sky localization of joint events, greatly decreasing the cost of follow-up searches with narrow field-of-view instruments. For example, IceCube sky location reconstruction for charged current $\nu_\mu$ interactions typically has an angular uncertainty $\lesssim 1^\circ$~\citep{PhysRevLett.124.051103}, much smaller than tens to hundreds, or even thousands, of square degrees that are normal for GW sky localizations~\citep{Abbott_2020prospects}. Consequently, joint multimessenger observations can constrict the sky localization of the observed events, making further follow-ups much more feasible.

The current wealth of multimessenger data provides the opportunity for statistically refined searches~\citep{2021ApJ...908..216V} aimed at identifying common cosmic sources of GWs and HENs. In this article, we describe a comprehensive archival search for HEN emission from GW candidates observed during the entirety of the Advanced LIGO and Advanced Virgo detectors' third observing run (O3), including sub-threshold event candidates, i.e. candidate events with higher false alarm rates than in catalogs. This search complements the previous searches and represents the most sensitive offline search done for joint emission of HEN and GW events, using a more extended dataset including data from a more sensitive run of the GW detectors. Having summarized the state of the field, we continue in Section~\ref{sec:NuDetector} where we describe the IceCube detector and the specific neutrino data that is used in our analysis. In Section~\ref{sec:GWDetectors}, we describe the GW detectors and the GW data. In Section~\ref{sec:Method}, we present the core concepts and distinctive aspects of our search methodology, implemented through the \gls{llama} analysis software~\citep{2019arXiv190105486C}, while also pointing to more intricate methodological discussions in previous works (e.g., ~\cite{2019PhRvD.100h3017B,2020ApJ...898L..10A,2021ApJ...908..216V}). Section~\ref{sec:Results} presents the outcomes of our search. Finally, in Section~\ref{sec:Conclusion}, we contextualize our findings and offer perspectives on future directions in this field.

\section{The IceCube HEN Detector and its Data} \label{sec:NuDetector}

IceCube is positioned at the geographic South Pole in Antarctica, utilizing a cubic kilometer of clear ice for its detection medium~\citep{2017JInst..12P3012A}. The IceCube Collaboration has deployed 5160 optical modules distributed across 86 vertical strings, to observe the depths of Antarctic ice between 1500 and 2500 meters. The main sensors within the optical modules are photomultiplier tubes designed to capture Cherenkov light emissions resulting from the interactions of neutrinos within the ice. 

IceCube neutrino events can be broadly categorized as \textit{tracks} and \textit{cascades}. \textit{Tracks} are events where muons or anti-muons traverse a linear trajectory, leaving a trail of detected Cherenkov light in their path. Atmospheric muons and atmospheric muon neutrinos are the major sources of background contamination for the set of astrophysical neutrino induced tracks events. \textit{Cascades} refer to showers occurring in the ice, originating from charged-current interactions of electron neutrinos or neutral-current interactions. These event topologies have different relevance for multimessenger searches depending on their different directional accuracy on the sky. For tracks, the typical reconstruction accuracy is $\lesssim \, 1^\circ$~\citep{PhysRevLett.124.051103}, whereas for cascades it is $\gtrsim \,10^\circ$. In this search, we use tracks-based neutrino triggers due to their better localization which is crucial for multimessenger astronomy.

IceCube maintains a duty cycle exceeding 99\% and can observe neutrinos from the entire sky~\citep{Pizzuto:2021RV}. IceCube data based on muon tracks consist primarily of background signals from muons and neutrinos originating in the Earth's atmosphere. The background in the Northern Hemisphere primarily comprises atmospheric neutrinos, while in the Southern Hemisphere, atmospheric muons predominate. As a result, the detector has less background in the Northern Hemisphere and the energy threshold for the detection of tracks in the Northern Hemisphere can be kept significantly lower ($\cal{O}$(1)\,TeV) than that in the Southern Hemisphere ($\cal{O}$(100)\,TeV).

The data analysis presented here relies on a low-latency (${\sim}30$~s) event reconstruction, facilitating timely multimessenger follow-up endeavors to IceCube events, known as the Gamma-ray Follow-Up (GFU) data stream~\citep{2016JInst..1111009I,Kintscher_2016,2019ICRC...36.1021B}. GFU events are well-reconstructed tracks, which enable precise sky localization essential for effective collaborative astrophysics with electromagnetic observatories. Although the low-latency property is not relevant for this article, the LLAMA pipeline has also been used in
real-time searches for coincident GW and HEN events, and in that case
the low latency is crucial~\citep{2019ICRC...36..930K}.

The analysis presented is based on a considerably larger number of candidate cosmic HEN events compared to the number of GW candidates, in average ${\sim}6.4$ neutrinos in the search time window of each candidate GW event.

\section{GW Detectors and their Data} \label{sec:GWDetectors}

The LIGO and Virgo detectors~\citep{2015CQGra..32g4001L,2015CQGra..32b4001A} are Michelson-type interferometers spanning multiple kilometers. LIGO has two nearly identical detectors, separated by 3000 km, situated in Hanford, Washington and in Livingston, Louisiana within the United States. Virgo operates a single detector located in Cascina, Italy. 

On April 1, 2019, the LIGO and Virgo detectors commenced their third observing run O3, which was divided into two periods. The initial part (O3a) encompassed the period from April to September 2019, while the second part (O3b) spanned November 2019 to March 2020. Over the course of O3, the GW detectors amassed datasets of unprecedented size and quality at the time~\citep{2020PhRvD.102f2003B,2021PhRvX..11b1053A,2023PhRvX..13d1039A,2023CQGra..40r5006A}, with the GW strain data now being publicly accessible~\citep{2023ApJS..267...29A}.

Not every IceCube GFU candidate coincided with GW detector observations as their duty factors were lower during O3 than IceCube's. For the first half of O3 they were 71\% for Hanford, 76\% for Livingston, and 76\% for Virgo; in the second half of the run 79\%, 79\%, and 76\% respectively~\citep{2021CQGra..38m5014D}.
The cosmic reach of the Hanford, Livingston, and Virgo detectors remained comparable during the first and second parts of the third observing run: 108 Mpc, 135 Mpc, and 45 Mpc of BNS inspiral range~\citep{PhysRevD.47.2198,Chen_2021} versus 115 Mpc, 133 Mpc, and 51 Mpc, respectively~\citep{2023ApJS..267...29A}.

The optimal joint treatment of events from IceCube, LIGO, and Virgo in the analysis relies on the localizations of both neutrinos and GWs detected by the network. Given the better accuracy of neutrino localization compared to GW localization, this approach significantly enhances the sky-localization potential for joint multimessenger candidates to $\cal{O}$(1)~deg$^{2}$, beyond the localization capabilities of the GW network alone ($\cal{O}$(10--1000)~deg$^{2}$), presenting new opportunities for electromagnetic observations in low-latency or using archival data.

The analysis presented relies on GW candidates, having a false alarm rate below two per day from GWTC-2.1~\citep{Zenodo-GWTC2.1} and GWTC-3.0~\citep{Zenodo-GWTC3}. Most of these events are \emph{sub-threshold} compared to the conventional thresholds for confident detections, such as having an estimated probability of astrophysical origin ($p_{\rm astro}$) greater than 0.5~\citep{2021arXiv210801045T,2023PhRvX..13d1039A}. 

We consider events from three template-based searches using compact binary coalescence (CBC) waveforms~\citep{2023PhRvX..13d1039A} commonly used by the LVK: GstLAL~\citep{2017PhRvD..95d2001M,2019arXiv190108580S,2020PhRvD.101b2003H,2021SoftX..1400680C}, Multi-Band Template
Analysis ~\citep[MBTA;][]{2016CQGra..33q5012A,2021CQGra..38i5004A}, and PyCBC~\citep{2012PhRvD..85l2006A,2005PhRvD..71f2001A,2014PhRvD..90h2004D,2016CQGra..33u5004U,2017ApJ...849..118N,2020PhRvD.102b2004D}. The latter was implemented in two versions, PyCBC-broad and PyCBC-BBH.

A total of 2210 GW candidates from CBC searches are used here in the joint GW+HEN analysis. This set includes the 79 CBC candidates confidently identified by at least one of the template-based searches with a $p_{\rm astro}$ greater than 0.5; of which $10$--$15\%$ may be contamination from triggers of terrestrial origin~\citep{2023PhRvX..13d1039A}. When a CBC trigger appeared in the candidate lists of multiple pipelines, the metadata (localization, $p_{\rm astro}$, etc.) associated with the pipeline having the highest probability of astrophysical origin was used in the joint analysis.

A total of 481 non-public candidates from the all-sky search for generic GW bursts from O3~\citep{2021PhRvD.104l2004A} were also selected for joint analyses. The selected triggers were produced by the coherent Wave  Burst (cWB) search pipeline~\citep{PhysRevD.93.042004,DRAGO2021100678} specifically designed to find candidates without explicit model prediction. The cWB pipeline, even in its generic all-sky mode, is capable of finding some of the CBC triggers~\citep{2021PhRvD.104l2004A,2023PhRvX..13d1039A}. We are using cWB-allsky triggers with selection criteria based on the correlation coefficient~\citep{PhysRevD.93.042004}, which quantifies the relative proportion of the coherent energy in the detector network's data, requiring it to be greater than 0.8 for Hanford--Livingston detections and greater than 0.5 for Hanford--Virgo and Livingston--Virgo detections~\citep{2021PhRvD.104l2004A}. For the cases of selected Hanford-Livingston candidates where Virgo data were also available, we produced the three-detector localizations dedicated for the analysis presented here, as source localization overlap is a deciding factor in multimessenger searches. 

\section{Methodology} \label{sec:Method}
Our analysis has two parts. First, we analyze the GW triggers one by one and find the HEN coincidence significance for each of them. Second, we evaluate them collectively to learn about the astrophysical population of jointly GW and HEN emitting sources.

\subsection{Analysis of individual events}

Previous studies of joint sources of GWs and HENs using sub-threshold trigger sets from both types of messengers during the initial GW detectors era and the first observing run of Advanced LIGO
~\citep{2008CQGra..25k4039A,2009IJMPD..18.1655V,PhysRevLett.107.251101,2012PhRvD..85j3004B,2013JCAP...06..008A,2013RvMP...85.1401A, 2014PhRvD..90j2002A,2019ApJ...870..134A} employed likelihood methods.
The analysis presented here employs the \gls{llama} pipeline~\citep{2019arXiv190105486C}, which was previously used for neutrino follow-up of \textit{confident} GW candidates~\citep{2020ApJ...898L..10A,2022icrc.confE.950V,2023ApJ...944...80A}.
The \gls{llama} pipeline uses an optimal model-dependent method~\citep{2019PhRvD.100h3017B} for GW+HEN searches.
LLAMA uses prior probabilities informed by astrophysics and detector characteristics to combine several hypotheses optimally for a physical model and reduces the problem to a simple test of two hypotheses. The odds ratio of having a multimessenger detection or not becomes the optimal test statistic for the physically motivated model.

Our priors are on the distribution of the source parameters $\boldsymbol \theta=\{D_\mathrm{L}, \mathbf{\Omega}_{s}, t_s, E_{\rm GW}, E_{\nu}\}$, which are the luminosity distance of the source $D_\mathrm{L}$ ($\propto D_\mathrm{L}^2$ prior distribution up to the GW detection range), its sky position $\mathbf{\Omega}_{s}$ (uniform prior distribution over the sky), the reference time for the astrophysical event $t_s$ (uniform distribution in the observing period), the isotropic equivalent emission energies in GWs $E_{\rm GW}$ and in HENs $E_{\nu}$. Beyond solving the optimal testing problem, the use of assumed emission energies also allows the use of the distance information from GW triggers optimally which further increases the statistical power of the search~\citep{2019PhRvD.100h3017B}. For GWs, the emission energies (log uniform distribution in $[10^{-1},10^{1}]~$M$_{\odot}c^2$) were chosen by considering the released energies in CBCs that have a total mass between 2~M$_\odot$ and 200~M$_\odot$. 

Short GRBs produced in BNS mergers are the most promising known high-energy emission mechanism for coincident GW and HEN detection, where the energy emitted in HENs is expected to be comparable with the energy emitted in gamma-rays~\citep{Kimura_2017,10.1093/mnras/sty285,2019ARNPS..69..477M}. Consequently, the HEN emission energies (log uniform distribution in $[10^{46},10^{51}]$~erg) were chosen according to an empirical distribution of short GRB photon emission energies~\citep{2014ARA&A..52...43B}. We assume the differential number density of the neutrino emission energy spectrum is a power law with exponent $-2$ which is expected for the neutrinos from cosmic rays accelerated by Fermi acceleration~\citep{1983RPPh...46..973D,Kurahashi_2022}.

Although the searches for HENs coincident with GRBs have not yielded any detection~\citep{2013RvMP...85.1401A,2017ApJ...849..153F,2018PhRvD..98d3020K}, a potential correlation between gravitational waves and high-energy neutrinos may still be explained through the GRB connection. As neutrino emission is less confined by beaming or opacity, a neutrino observed in coincidence with a neutron star merger would indicate off-axis, electromagnetically obscured, or failed GRB source.

Our observational inputs ($\mathbf{x}$) from the GW side are the localization, candidate event detection time, signal-to-noise ratio ($\rho$), and for template-based searches source distance information and $p_{\rm astro}$.
From the neutrino detector side, the inputs are the reconstructed time of arrival, energy proxy (related to the energy deposited in the detector), direction, and angular uncertainty of the HENs.

The standard prompt search time window is $\pm500$~s and was established on the assumption of a GRB source model~\citep{2011APh....35....1B}. Our signal likelihood is maximum when there is no time difference between the neutrino and the GW candidate, and it decreases to zero linearly as the time difference approaches $\pm500$~s. We evaluate all of the neutrino candidates coincident with the GW trigger within this time window together.

Our analysis is based on testing a signal hypothesis $H_{s}$: the GW and at least one HEN come from the same astrophysical source, against multiple background hypotheses: both the GW and HENs are noise originated ($H_{n}$), the GW is astrophysical while all HENs are noise originated ($H_{c}^{\rm GW}$), i.e.\ atmospheric muons or neutrinos, and finally the GW is noise originated and a HEN is astrophysical ($H_{c}^{\nu }$). Other very low probability occurrences (about 2 orders of magnitude less probable than the aforementioned cases) which include more than one astrophysical but unrelated messengers (i.e.\ a GW coming from an astrophysical event and a neutrino coming from a separate astrophysical event) are ignored. In the case of HENs, noise refers to atmospheric muons or neutrinos.
We combine the background hypotheses by using their prior probabilities which were calculated using our assumed source parameters and detector characteristics. Our test statistic (TS) is 
\begin{widetext}
\begin{equation}
\label{TS}
{\rm TS}(\mathbf{x})=\frac{P(\mathbf{x}|H_s)P(H_s)}{P(\mathbf{x}|H_n)P(H_n)+P(\mathbf{x}|H_c^{\rm GW})P(H_c^{\rm GW})+P(\mathbf{x}|H_c^{\nu})P(H_c^{\nu})}.
\end{equation}
\end{widetext}
The prior probability of $H_s$ is an overall constant in the TS and does not affect any of our results.

The analysis presented here differs from previous analyses with \gls{llama}~~\citep{2020ApJ...898L..10A,2022icrc.confE.950V,2023ApJ...944...80A} where we used only confident detections of GWs (e.g., $p_{\rm astro}>0.5$) and thus the first and third terms in the denominator of Eq. \eqref{TS} were taken as zero. The original analysis method in \cite{2019PhRvD.100h3017B} is suitable for the analyses with confident or sub-threshold GW triggers, where only $\rho$ was used to estimate the signalness of the GW candidates. Here we improve it for the triggers of the template-based searches by using an additional input: $p_{\rm terr}=1-p_{\rm astro}$ which is the estimated probability of terrestrial origin for the GW triggers. For these triggers, we modify the way we calculate the joint probability density of noise origin for GWs and signal-to-noise ratio as 
\begin{widetext}
    
\begin{multline}
P(\rho,{\rm GW\ noise})=P(\rho)P({\rm GW\ noise}|\rho)= \left[\frac{P(\rho|{\rm GW\ noise})R_{\rm bgGW}+3\rho_{\rm min}^3/\rho^{4} (4\pi D_\mathrm{L-max}^3 \dot{n}_{\rm GW\nu}/3)}{R_{\rm bgGW}+4\pi D_\mathrm{L-max}^3 \dot{n}_{\rm GW\nu}/3}\right] \times p_{\rm terr}.
\end{multline}
\end{widetext}
Here we used $P({\rm GW\ noise}|\rho)=p_{\rm terr}$. The expression in the square brackets is $P(\rho)$. Here $R_{\rm bgGW}$ is the rate of GW background triggers, $ \dot{n}_{\rm GW\nu} $ is the rate density of astrophysical multimessenger emitting events and $D_\mathrm{L-max}$ is the horizon distance of the GW detectors for the minimum signal-to-noise ratio in our dataset $\rho_{\rm min}\approx5.6$. We take $ \dot{n}_{\rm GW\nu} $ constant as 23.9~Gpc$^{-3}$yr$^{-1}$, which is the most recent estimate of the local rate density of BBH mergers \citep{PhysRevX.13.011048} that make up the most of the detections. For this calculation only, we ignore the cosmological expansion for simplicity, which may have a minuscule negative effect on the statistical power of the search; but does not have a direct effect on our physical inferences otherwise. The probability $P(\rho)$ is estimated by using the conditional distributions of $\rho$ when the GW candidate is of noise and astrophysical origin; and the two scenarios' expected relative frequencies. The conditional distributions are $P(\rho|{\rm GW\ noise})$ and $3\rho_{\rm min}^3\rho^{-4}$ respectively for noise and astrophysical origins. While the former is estimated empirically, the latter can be derived assuming homogeneously distributed sources that are detected according to a $\rho$ threshold~\citep{Schutz_2011}. The relative frequencies are proportional to $R_{\rm bgGW}$ and $(4/3)\pi D_\mathrm{L-max}^3 \dot{n}_{\rm GW\nu}$.

In order to evaluate the significance of each joint candidate, we use the Bayesian odds ratio in Eq.~\eqref{TS} as a test statistic. We find the frequentist significance by comparing the odds ratio for each event to a distribution built empirically using background data and simulations. Coincidences are simulated by randomly matching neutrinos from the GFU stream and the GW trigger events we analyze at the detection rate of the GFU stream (6.4 neutrino triggers per GW trigger on average in 1000~s of search window). We use different background distributions for CBC and cWB triggers.

\subsection{Population analysis}

The individual analysis described above can uncover individual events that become high-significance once GW and neutrino information is combined. We carried out a separate population analysis that combines significances from multiple events and aims to find an excess significance from the whole dataset to examine the presence of a joint emission from \emph{any} of the events. In this case, even if no individual event is sufficiently significant to claim a discovery, we can statistically infer the presence of a signal by finding that the significance distribution shows an excess of high-significance events compared to what would be expected from background.

For our population study we constructed a new test statistic ${\rm TS_{pop}}$ (Eq. \eqref{eq:TSpop}); because the new question we would like to answer now is whether there are any detections in the dataset or not. When the \emph{a priori} expected number of multimessenger events in the dataset is low, the optimal test statistic is equivalent to the \textit{sum} of the individual test statistics from the individual observations. Hence, we construct ${\rm TS_{pop}}$ as the sum of the test statisics of the analyzed events. Here, a single event should be understood as the combination of neutrinos within the search time window with a GW candidate. The sum is over all such combinations in the whole observation run
\begin{equation}
    {\rm TS_{pop}}=\sum_i {\rm TS}(\mathbf{x}_i).
    \label{eq:TSpop}
\end{equation}

While such a population study does not identify specific joint events as significant, it has more power to test whether there \textit{are} such events in the dataset. It can be thought of as a more efficient trials correction which uses all the information in the whole dataset. 

The $p$-value of this analysis was found by constructing an appropriate background distribution. This was done by repeatedly sampling events from our previously generated background events as many as the analyzed events (2210 for CBC and 481 for cWB) and summing their TSs.

In order to constrain the rate density of joint GW+HEN emitters, we need to characterize the behaviour of the new test statistic when there are different numbers of multimessenger events in the dataset. For this, we injected simulated multimessenger signals to the background event set and created ${\rm TS_{pop}}$ distributions with different number of (0, 1, 2, ...) simulated multimessenger events. These essentially differ from the 0 multimessenger event case by the individual TS of the extra multimessenger events. In order to create such multimessenger events we artificially injected a GW signal and a neutrino from the same source location. For this we simulate GW events and combine them with neutrinos.

The additional sub-threshold GW events were generated by injection simulations using BAYESTAR~\citep{2016PhRvD..93b4013S,Singer_2016}. We simulated GW events using the properties of the LIGO and Virgo detectors and taking into account the properties of the observed populations for each detector combination during O3 \citep{2023PhRvX..13d1039A}. For each source type of BNS, neutron star--black hole binaries (NSBH), and BBH, we performed randomized injections for the corresponding source populations' parameter (mass, spin, sky position, inclination, distance) distributions, calculated the signal-to-noise ratio, and generated the corresponding localizations considering a network of at least two detectors. The \textsc{POWERLAW+PEAK} model~\citep{Talbot_2018} with the median estimations of the parameters from the analysis of GWTC-3.0~\citep{PhysRevX.13.011048} was used for the mass distribution of the injected BBH mergers. For neutron stars, we used 1.4~M$_\odot$. With the insufficient observational preference for the mass distribution of NSBH, we simply drew the black hole masses from the primary mass distribution of the BBH mergers and the neutron star masses as 1.4~M$_{\odot}$. For the simulated GW events we assigned $p_{\rm astro}$ values by using the public injection set for GW detection pipelines~\citep{Zenodo-sensitivity}. We found the injection in that dataset that has the most similar parameters to the parameters of our injections, for each of our injections, and assigned the same $p_{\rm astro}$ value. This is done by defining a unitless statistic which quantifies the similarity of the events based on the product of the differences in the parameters ($\chi_{\rm eff}$ spin parameter, chirp mass, luminosity distance, and the sky position in the detector frame) scaled by the expected measurement uncertainties.

The background neutrinos for each injected GW signal are drawn from the scrambled GFU sample. The 90\% central energy range of the GFU sample ranges from $500$ GeV$-50$ PeV, assuming a source with $\epsilon^{-2}$ spectrum. This simple spectrum can be viewed as a nominal spectrum without the complicated different structures predicted in theoretical studies, and we generate signal neutrinos assuming this spectrum for the differential number density. We inject one Monte Carlo generated signal neutrino corresponding to the injection sky localization position of the signal GW event.

The population upper limit at a confidence level $CL$ then can be found by the standard Neyman's construction~\citep{Neyman:1937uhy} by requiring the probability of having ${\rm TS_{pop}}$ for the same sample size containing injected multimessenger signal(s) being higher than the observed ${\rm TS_{pop}}$ to be $CL$ or higher.

The upper limit on the multimessenger (MM) rate density $\dot{n}_{\rm GW\nu}$, depending on the neutrino emission energy $E_\nu$ and the GW source type ($\mathcal{S}$, e.g., BNS, NSBH, BBH) can be calculated as

\begin{widetext}
  \begin{multline}
P({\rm TS_{pop}}>{\rm TS^{observed}_{pop}}|E_{\nu}, \mathcal{S}, \dot{n}_{\rm GW\nu})
= \sum_{\#_{\rm MM}=0}^{\substack{{\rm No\ of\ obs.}\\{\rm GW\ events}}}  P({\rm TS_{pop}}>{\rm TS^{observed}_{pop}}| \mathcal{S}, \#_{\rm MM}) P(\#_{\rm MM}| E_{\nu}, \mathcal{S}, \dot{n}_{\rm GW\nu}) = {\rm CL},
  \end{multline}
where the sum is over the MM event count in the dataset ($\#_{\rm MM}$) whose occurrence probability is
  \begin{multline}
P(\#_{\rm MM}| E_{\nu}, \mathcal{S}, \dot{n}_{\rm GW\nu})
= {\rm Poisson}\left( \#_{\rm MM}, {\rm Mean}=\dot{n}_{\rm GW\nu}\times T_{\rm obs}\int 2\pi f(D_\mathrm{L},z) p^\nu_{\rm det}(E_{\nu},D_\mathrm{L},\delta)p^{\rm GW}_{\rm det}(\mathcal{S},D_\mathrm{L})\cos\delta {\rm d}\delta {\rm d}D_\mathrm{L}\right),
  \end{multline}
\end{widetext}
where $p^\nu_{\rm det}(E_{\nu},D_\mathrm{L},\delta)$ and $p^{\rm GW}_{\rm det}(\mathcal{S},D_\mathrm{L})$ are the detection probabilities of neutrinos and GWs respectively from an event with total neutrino emission energy $E_{\nu}$ and the specific GW event type $\mathcal{S}$, as a function of luminosity distance $D_\mathrm{L}$, and declination ($\delta$). The total observation time is represented by $T_{\rm obs}$, ${\rm Poisson}(a,{\rm Mean}=b)=b^ae^{-b}/a!$ is the probability of a Poisson point process for $a$ happened events with mean $b$, while $f(D_\mathrm{L},z)$ gives the distribution of MM events as a function of distance and redshift $z$. Considering the cosmological expansion,
\begin{multline}
    f(D_\mathrm{L},z)=(1+z)^{-1}\frac{{\rm d}V_{\rm c}}{{\rm d}D_\mathrm{L}}\\=\frac{c}{(1+z)^3H_0\sqrt{\Omega_m(1+z)^3+\Omega_\Lambda}}\frac{{\rm d}z}{{\rm d}D_\mathrm{L}}D_\mathrm{L}^2,
    \label{eq:cosmo}
\end{multline}
assuming a constant rate in comoving volume $V_{\rm c}$ where $c$ is the speed of light, $H_0=67.9~$km s$^{-1}$ Mpc$^{-1}$ is the Hubble constant, $\Omega_{\rm m}=0.3065$ and $\Omega_\Lambda=0.6935$ are the energy densities of matter and cosmological constant~\citep{2016A&A...594A..13P}. We find $p^{\rm GW}_{\rm det}(\mathcal{S},D_\mathrm{L})$ by calculating the detection fraction of our injected GW events as a function of their distance for different source types, averaged over their remaining properties including their sky positions. The sky averaging is a sensible approximation as the GW detection probability would have a mild dependency on the sky location after averaging only over the Earth's rotational motion. On the other hand, the detection probability of neutrinos has a strong dependency on the declination after being averaged only over the Earth's rotational motion. Therefore its declination dependent form was used, which is the Poisson probability of observing at least one neutrino. The mean number of detected neutrinos for this Poisson probability is expressible in terms of the effective area of the detector $A_{\rm eff}$ and the isotropically equivalent emitted energy in neutrinos ($E_{\rm iso}$) as
\begin{equation}
    \langle n_\nu \rangle(E_{\rm iso},D_\mathrm{L},\delta)=\frac{1}{3}\frac{E_{\rm iso}}{4\pi D_\mathrm{L}^2 \ln(\epsilon_{\rm max}/\epsilon_{\rm min})}\int \frac{A_{\rm eff}(\epsilon,\delta)}{\epsilon^2}{\rm d}\epsilon,
    \label{eq:nnu}
\end{equation}
where the ratio of the limits of the relevant energy range is $\epsilon_{\rm max}/\epsilon_{\rm min}\approx10^6$, the factor 1/3 accounts for the fact that only a third of the astrophysical HEN are expected to be muon neutrinos \citep{ATHAR_2006,Pakvasa_2008} and we assumed an $\epsilon^{-2}$ spectrum for the emitted neutrino number density~\citep[see also Eq. 3 in][]{2020ApJ...898L..10A}. For isotropic emissions, isotropically equivalent emission energy is equal to the total emitted energy: $E_{\rm iso}=E_\nu$. For beamed emissions with beaming factor $f_{\rm beam}$, $E_{\rm iso}=f_{\rm beam}E_\nu$ within the emission cone and zero outside. With these we can write the probability of detecting neutrinos as
\begin{equation}
    p^\nu_{\rm det}(E_{\nu},D_\mathrm{L},\delta)=1-\exp(-\langle n_\nu \rangle(E_{\nu},D_\mathrm{L},\delta)).
\end{equation}

\section{Results} \label{sec:Results}

\subsection{Individual p-values}

There are on average 6.4 GFU track neutrinos within any 1000~s time segment in our data. Of the $2210$ CBC candidate events, 5 did not have coincident neutrinos within the $\pm$500~s search time window around the GW trigger time, resulting in individual $p$-values of 1 for each of these events. Altogether 14491 time coincident neutrino tracks were identified and analyzed under a joint source hypothesis; none of them had a sufficiently high significance to claim a detection considering the number of analyzed GW triggers. 

\begin{figure}
    \includegraphics[width=\columnwidth]{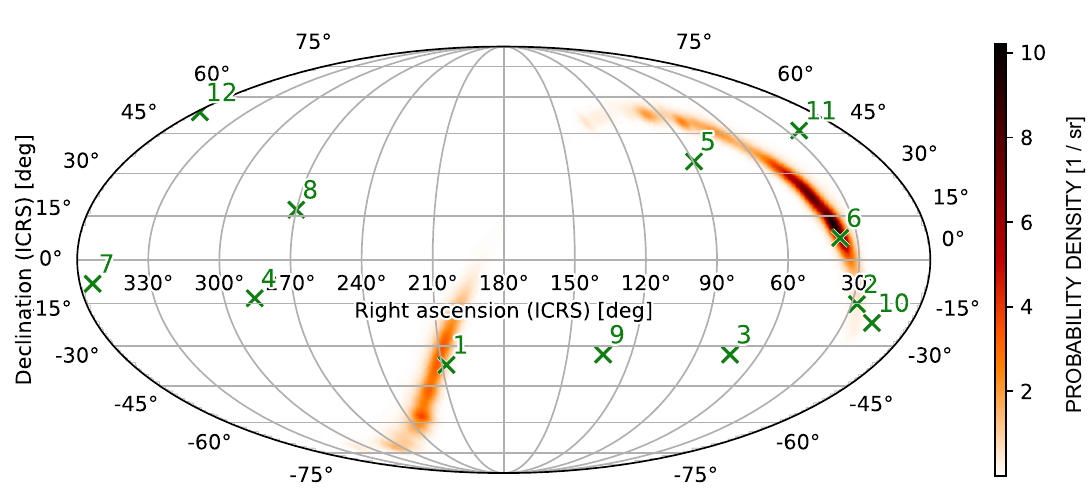}
    \caption{Joint sky localization showing all neutrino candidates coincident with GW BNS candidate at GPS time~=~1262142545.615. Neutrino~\#6 gives the dominant contribution to the significance.}
    \label{fig:bestCBCresult}
\end{figure}

The lowest $p$-value candidate, presented in Fig.~\ref{fig:bestCBCresult}, has an individual $p$-value of $3.8\times10^{-4}$ (0.84 post-trials after multiplying by the number of analyzed events). The GW trigger is a BNS candidate; but it is much more likely to be of terrestrial origin ($p_{\rm terr}=0.997$). The coincident neutrino number 6, which produces most of the significance, has a reconstructed energy of 2.4~TeV. Its sky position is right above the equator, corresponding to the high sensitivity declination region of the IceCube detector, while its energy is not particularly outstanding from the expected background. The reconstructed mean localization distance for the GW candidate's source is 295~Mpc. The GW trigger occured 222~s after the arrival time of neutrino number 6.

\begin{figure}
    \includegraphics[width=\columnwidth]{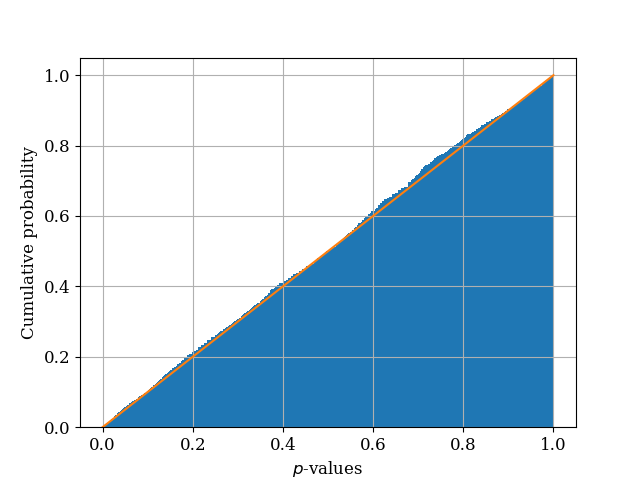}
    \caption{Cumulative distribution of the $p$-values from the CBC part of the analysis. The orange line is a reference for uniform distribution.}
    \label{fig:CBCpvaldist}
\end{figure}

The cumulative distribution of all of the $p$-values for CBC candidates is shown in Fig~\ref{fig:CBCpvaldist}. It is consistent with a uniform $p$-value distribution, which is expected in the absence of any joint emission. The Kolmogorov--Smirnov test $p$-value when comparing to a uniform distribution is 0.05, which is consistent with the expectation from background.

\begin{figure}
    \includegraphics[width=\columnwidth]{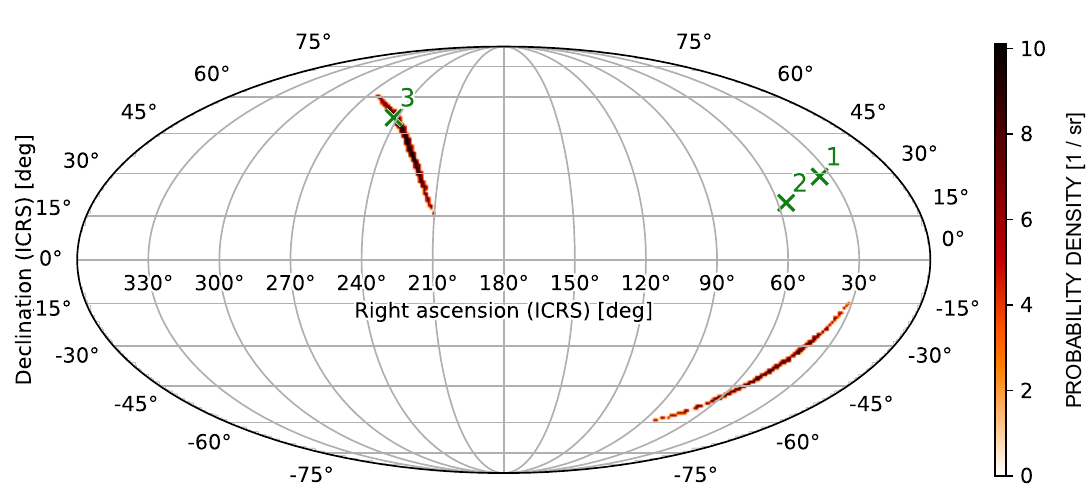}
    \caption{Joint sky localization showing all neutrino candidates coincident with GW burst candidate at GPS time~=~1241247887.938. Neutrino~\#3 gives the dominant contribution to the significance.}
    \label{fig:cwbmost}
\end{figure}

For cWB events, we similarly did not have a sufficiently high-significance case to claim a joint detection considering the number of analyzed triggers. The lowest $p$-value candidate, with an individual $p$-value of 4.3$\times10^{-4}$ (0.21 post-trials after multiplying by the number of analyzed events), is shown in Fig. \ref{fig:cwbmost}. The coincident neutrino number 3, which produces most of the significance, has a reconstructed energy of 2.3~TeV. It was detected 154~s after the GW trigger.

\begin{figure}
    \includegraphics[width=\columnwidth]{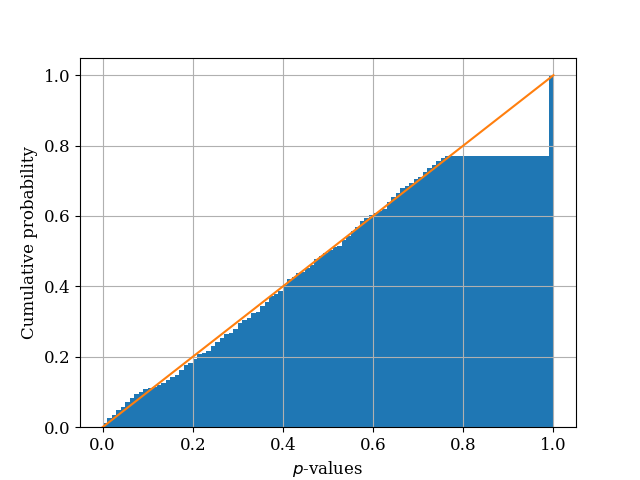}
    \caption{Cumulative distribution of the $p$-values from the cWB part of the analysis. The orange line is a reference for uniform distribution. The discrete behavior at the right side is due to cWB sky localizations not having smooth probability distributions, most of the sky possessing exactly zero probability.}
    \label{fig:Cwbpvaldist}
\end{figure}

The cumulative distribution of all of the $p$-values of the cWB part of the analysis is shown in Fig~\ref{fig:Cwbpvaldist}. It is consistent with a uniform $p$-value distribution with a discrete behaviour at high $p$-values. This discrete behaviour is caused by cWB sky localizations not containing low probability densities below a certain threshold, which causes non-smooth distributions. The Kolmogorov--Smirnov test $p$-value is 0.67 in comparison with a uniform distribution with a similar discrete behaviour.

We provide the properties of the neutrino triggers that produce individual $p$-values of 0.01 or less in Table~\ref{tab:nutable} in the Appendix.
\subsection{Population Constraints}

When the whole dataset was analyzed collectively, we obtained a $p$-value of 0.22 for CBC triggers and 0.18 for cWB triggers. Consequently, we conclude that there is no significant sign of joint detection of GW and HEN in our dataset. We obtained frequentist 90\% upper limits for the rate density of multimessenger events related to CBC triggers, assuming a homogeneous distribution in comoving volume. HEN emission from CBC sources is expected to be beamed. We calculate different limits assuming an isotropic neutrino emission and a beamed emission with a beaming factor of $f_{\rm beam}\sim100$~\citep{Fong_2015}. We find the limits for different time integrated bolometric neutrino emission luminosities (total $E_\nu$) which are assumed to be the same for all GW events. For the beamed emission we assume the total energy to be concentrated in a solid angle of $4\pi/f_{\rm beam}$, directed isotropically without considering correlations with the GW emission geometry. This is a conservative assumption as the beamed jets are predicted to be directed along the orbital axis of the binaries along which the GW emission is also more powerful. 

\begin{figure}
    \includegraphics[width=\columnwidth]{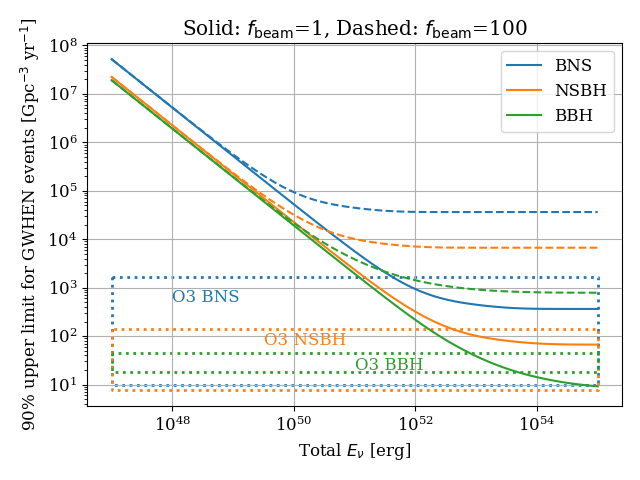}
    \caption{90\% upper limits on GW+HEN source population for different CBC source types as a function of the total neutrino emission energy from each source. We assume identical neutrino emission characteristics for sources with an $\epsilon^{-2}$ spectrum. The GW related properties are according to the injections which were guided by the O3 inferences~\citep{PhysRevX.13.011048}. The upper limits are shown for isotropic neutrino emission with solid lines and for a beamed emission with a beaming factor of 100 with the dashed lines. The rectangular regions described by dotted lines correspond to the estimated rate densities of the corresponding GW sources after O3 with 90\% credibility~\citep{PhysRevX.13.011048}.}
    \label{fig:popUL}
\end{figure}

The limits on GW+HEN source population are shown in Fig.~\ref{fig:popUL} together with the estimated 90\% credible intervals of the rate densities of different CBC sources~\citep{PhysRevX.13.011048}. We show the BBH rate density at $z=0.2$, based on a model in which the BBH rate density evolves as $(1+z)^\kappa$ with an estimated $\kappa=2.9^{+1.7}_{-1.8}$. For any of our GWHEN limits we do not assume such a redshift evolution. Since the observed sources are all below $z\lesssim1$, this modeling difference can bring at most $\mathcal{O}(1)$ variations to the compared BBH rate, which would not change our conclusions substantially. We can constrain any HEN emission from a GW source if our derived upper limits lie below the estimated rate density of GW sources. However, the upper limits at 90\% confidence level and the 90\% credible intervals can only be compared with the caveat that one is a frequentist concept and the other a Bayesian one. The 90\% confidence level population upper limits assume neutrino emission with an $\epsilon^{-2}$ spectrum between $100$~GeV and $100$~PeV. The energy bounds of the spectrum were chosen according to the sensitivity of IceCube to track events as in previous analyses~\citep{2020ApJ...898L..10A,2023ApJ...944...80A}. Although these bounds can vary from model to model, the multimessenger upper limits weakly depend on these bounds as they enter the calculation only logarithmically (Eq.~\eqref{eq:nnu}) for an $\epsilon^{-2}$ spectrum. 

Even our most stringent joint source population upper limit shows that neutrino emission from GW sources can only be constrained at very high neutrino emission energies (time integrated bolometric luminosity~$>10^{52}$--$10^{54}$~erg) with isotropic emission. These energies can be compared with the beaming-corrected most energetic short GRB energies of $\sim10^{51}$~erg~\citep{2014ARA&A..52...43B}. For beamed neutrino emissions or lower total neutrino emission energies, which correspond to typical GRB emission characteristics, the upper limits on the joint source rate densities lie above the estimated rate densities of GW sources; hence they do not constrain the joint emission from GW sources. The limits of beamed emissions remain higher compared to isotropic emission at high neutrino emission energies as the fact that there is a limited solid angle of emission also bounds the chance of neutrino observation independent of the emission strength. The values 90\% upper limits converge to at the higher end of neutrino emission energies are set by the detection capabilities of GW detectors; since a sufficiently high number of neutrinos can be detected from such energetic emissions and neutrino detection from a joint emission would not be constraining. That is also mostly the reason why limits for BBH mergers reaches lower values compared to BNS and NSBH mergers, as BBH mergers are detected by LIGO and Virgo much more easily due to the higher amplitude of emitted GW waves in the detectors' sensitive frequency band. On the other hand, for lower neutrino emission energies, the upper limits are set by the neutrino detection capabilities. These upper limits present the first constraints on the high-energy neutrino emission from the population of CBCs. Previous constraints have been put using only the candidate GW bursts which were from the first observing run of Advanced LIGO~\citep{2019ApJ...870..134A}.

In the context of the short GRB source population, we note that the rate of observable (on-axis) GRBs is estimated to be around 1\% of the O3 BNS rate shown on Fig.~\ref{fig:popUL}. While no high-energy neutrinos were observed in coincidence with a detected GRB, there is still room for observable neutrino emission from neutron star containing binary mergers through off-axis emission or from electromagnetically opaque environments. The GW+HEN population limits presented here started to chart that territory.

\section{Conclusion and Outlook} \label{sec:Conclusion}

The discovery and multimessenger observation of cosmic sources involving HENs and GWs hold the potential to shed light on the origin of the highest energy neutrinos and cosmic rays, reveal high-energy emission from GW engines, enable more effective electromagnetic follow-up efforts, and enhance our understanding of source dynamics. A detailed examination of the O3 data and coincident IceCube event candidates has significantly expanded the analyzed dataset, thereby facilitating a new search.

The \gls{llama} analysis, employing a model-dependent statistically optimal approach, was applied to all sub-threshold GW candidates within the O3 catalogs and all time-coincident IceCube event candidates. We searched for neutrino emission within a $\pm$500~s time window centered around the GW merger time. No significant neutrino emission was observed during the search for individual GW candidates. The intersections of the coinciding sky localizations showed the potential effectiveness of multimessenger searches in guiding the observing strategies of electromagnetic observatories by significantly reducing the area that needs to be scanned by telescopes.

The absence of a GWHEN definitive detection resulted in constraints on the population of cosmic multimessenger sources, involving both GWs and neutrinos. These limits can constrain the fraction of HEN emitting GW sources only at very high neutrino emission energies and for isotropic emission. This implies that the joint detection of GW and HEN is mainly limited by the neutrino detection capabilities, providing additional motivations for the next generation of neutrino detectors. We expect enhanced chances for the joint detection of GW and HEN, and improvements for such analyses with the development of next generation neutrino detectors.

The fourth observing run of LVK (O4) is ongoing, with an improved sensitivity~\citep{2025CQGra..42h5016S,2025ApOpt..64.4710V}. Beyond the confirmed 128 CBC mergers identified with $p_{\rm astro}\geq0.5$ and thousands of subthreshold candidates from the first portion of O4~\citep{2025arXiv250818082T}, there are already additional hundreds of significant mergers and subthreshold candidates to investigate for joint emission with neutrinos that were identified in low-latency~\citep{gracedb_latest,2025arXiv250818082T}. Real-time multimessenger searches involving HENs and GWs are currently operational on both confident and sub-threshold GW candidates collected during the fourth observing run. The successful operation of O4~\citep{2025arXiv250818080T} showcases improved performance, consequently expanding the scope of investigated compact object mergers. The growing network of gravitational-wave detectors will give rise to a heightened rate of GW exploration and more confined source localizations, which in turn will provide more opportunities for conducting multimessenger studies, increasing the likelihood of discovering common sources of neutrinos and GWs.

Next-generation ground-based GW detectors Einstein Telescope and Cosmic Explorer are currently in their design phase, with the goal of significantly expanding the cosmic volume reach by orders of magnitude~\citep{2019BAAS...51g..35R,2011arXiv1108.1423S,ETDesign,2019BAAS...51c.239K}. The planned space-based GW detectors LISA~\citep{DANZMANN20001129,amaroseoane2017laserinterferometerspaceantenna}, TianQin~\citep{Luo_2016} and Taiji~\citep{Ruan_2020} will probe the millihertz frequency regimes. Simultaneously, the neutrino detectors are poised to increase their instrumented volumes with the next generation detector projects IceCube-Gen2 and KM3NeT~\citep{2021JPhG...48f0501A, Biagi_2012}. Anticipating a substantial increase in both the rate and quality of observations from these future detectors, we look forward to heightened chances and excitement surrounding future multimessenger discoveries~\citep{2013RvMP...85.1401A,2017muas.book.....B,2010JPhCS.243a2001M,2019ARNPS..69..477M,2012ANYAS,2019NatRP...1..585M}.

\section*{Data availability}
Supplementing data to this article can be found at \url{https://dataverse.harvard.edu/dataverse/icecube}.
\section*{Acknowledgements} 
This work has made use of the following software in alphabetical order: \texttt{astropy}~\citep{astropy2022astropy}, \texttt{ipython}~\citep{perez2007ipython}, \texttt{LLAMA}~\citep{2019arXiv190105486C}, \texttt{matplotlib}~\citep{hunter2007matplotlib}, \texttt{numpy}~\citep{harris2020array}, \texttt{python}~\citep{python}, and \texttt{scipy}~\citep{virtanen2020scipy}.

The IceCube Collaboration acknowledges the significant contributions to this manuscript from Zsuzsa M\'arka, Do\u{g}a Veske and Albert Zhang. The authors gratefully acknowledge the support from the following agencies and institutions: USA – U.S. National Science Foundation-Office of Polar Programs, U.S. National Science Foundation-Physics Division, U.S. National Science Foundation-EPSCoR, U.S. National Science Foundation-Office of Advanced Cyberinfrastructure, Wisconsin Alumni Research Foundation, Center for High Throughput Computing (CHTC) at the University of Wisconsin–Madison, Open Science Grid (OSG), Partnership to Advance Throughput Computing (PATh), Advanced Cyberinfrastructure Coordination Ecosystem: Services \& Support (ACCESS), Frontera and Ranch computing project at the Texas Advanced Computing Center, U.S. Department of Energy-National Energy Research Scientific Computing Center, Particle astrophysics research computing center at the University of Maryland, Institute for Cyber-Enabled Research at Michigan State University, Astroparticle physics computational facility at Marquette University, NVIDIA Corporation, and Google Cloud Platform; Belgium – Funds for Scientific Research (FRS-FNRS and FWO), FWO Odysseus and Big Science programmes, and Belgian Federal Science Policy Office (Belspo); Germany – Bundesministerium für Forschung, Technologie und Raumfahrt (BMFTR), Deutsche Forschungsgemeinschaft (DFG), Helmholtz Alliance for Astroparticle Physics (HAP), Initiative and Networking Fund of the Helmholtz Association, Deutsches Elektronen Synchrotron (DESY), and High Performance Computing cluster of the RWTH Aachen; Sweden – Swedish Research Council, Swedish Polar Research Secretariat, Swedish National Infrastructure for Computing (SNIC), and Knut and Alice Wallenberg Foundation; European Union – EGI Advanced Computing for research; Australia – Australian Research Council; Canada – Natural Sciences and Engineering Research Council of Canada, Calcul Québec, Compute Ontario, Canada Foundation for Innovation, WestGrid, and Digital Research Alliance of Canada; Denmark – Villum Fonden, Carlsberg Foundation, and European Commission; New Zealand – Marsden Fund; Japan – Japan Society for Promotion of Science (JSPS) and Institute for Global Prominent Research (IGPR) of Chiba University; Korea – National Research Foundation of Korea (NRF); Switzerland – Swiss National Science Foundation (SNSF).

\bigskip
This material is based upon work supported by NSF’s LIGO Laboratory which is a major facility
fully funded by the National Science Foundation.
The authors also gratefully acknowledge the support of
the Science and Technology Facilities Council (STFC) of the
United Kingdom, the Max-Planck-Society (MPS), and the State of
Niedersachsen/Germany for support of the construction of Advanced LIGO 
and construction and operation of the GEO\,600 detector. 
Additional support for Advanced LIGO was provided by the Australian Research Council.
The authors gratefully acknowledge the Italian Istituto Nazionale di Fisica Nucleare (INFN),  
the French Centre National de la Recherche Scientifique (CNRS) and
the Netherlands Organization for Scientific Research (NWO), 
for the construction and operation of the Virgo detector
and the creation and support  of the EGO consortium. 
The authors also gratefully acknowledge research support from these agencies as well as by 
the Council of Scientific and Industrial Research of India, 
the Department of Science and Technology, India,
the Science \& Engineering Research Board (SERB), India,
the Ministry of Human Resource Development, India,
the Spanish Agencia Estatal de Investigaci\'on (AEI),
the Spanish Ministerio de Ciencia e Innovaci\'on and Ministerio de Universidades,
the Conselleria de Fons Europeus, Universitat i Cultura and the Direcci\'o General de Pol\'{\i}tica Universitaria i Recerca del Govern de les Illes Balears,
the Conselleria d'Innovaci\'o, Universitats, Ci\`encia i Societat Digital de la Generalitat Valenciana and
the CERCA Programme Generalitat de Catalunya, Spain,
the National Science Centre of Poland and the European Union – European Regional Development Fund; Foundation for Polish Science (FNP),
the Swiss National Science Foundation (SNSF),
the Russian Foundation for Basic Research, 
the Russian Science Foundation,
the European Commission,
the European Social Funds (ESF),
the European Regional Development Funds (ERDF),
the Royal Society, 
the Scottish Funding Council, 
the Scottish Universities Physics Alliance, 
the Hungarian Scientific Research Fund (OTKA),
the French Lyon Institute of Origins (LIO),
the Belgian Fonds de la Recherche Scientifique (FRS-FNRS), 
Actions de Recherche Concertées (ARC) and
Fonds Wetenschappelijk Onderzoek – Vlaanderen (FWO), Belgium,
the Paris \^{I}le-de-France Region, 
the National Research, Development and Innovation Office Hungary (NKFIH), 
the National Research Foundation of Korea,
the Natural Science and Engineering Research Council Canada,
Canadian Foundation for Innovation (CFI),
the Brazilian Ministry of Science, Technology, and Innovations,
the International Center for Theoretical Physics South American Institute for Fundamental Research (ICTP-SAIFR), 
the Research Grants Council of Hong Kong,
the National Natural Science Foundation of China (NSFC),
the Leverhulme Trust, 
the Research Corporation,
the National Science and Technology Council (NSTC), Taiwan,
the United States Department of Energy,
and
the Kavli Foundation.
The authors gratefully acknowledge the support of the NSF, STFC, INFN and CNRS for provision of computational resources.

This work was supported by MEXT, the JSPS Leading-edge Research Infrastructure Program, JSPS Grant-in-Aid for Specially Promoted Research 26000005, JSPS Grant-in-Aid for Scientific Research on Innovative Areas 2402: 24103006, 24103005, and 2905: JP17H06358, JP17H06361 and JP17H06364, JSPS Core-to-Core Program A.\ Advanced Research Networks, JSPS Grants-in-Aid for Scientific Research (S) 17H06133 and 20H05639, JSPS Grant-in-Aid for Transformative Research Areas (A) 20A203: JP20H05854, the joint research program of the Institute for Cosmic Ray Research, University of Tokyo, the National Research Foundation (NRF), the Computing Infrastructure Project of the Global Science experimental Data hub Center (GSDC) at KISTI, the Korea Astronomy and Space Science Institute (KASI), the Ministry of Science and ICT (MSIT) in Korea, Academia Sinica (AS), the AS Grid Center (ASGC) and the National Science and Technology Council (NSTC) in Taiwan under grants including the Science Vanguard Research Program, the Advanced Technology Center (ATC) of NAOJ, and the Mechanical Engineering Center of KEK.

\bibliography{References}{}

@ARTICLE{2016PhRvD..93b4013S,
       author = {{Singer}, Leo P. and {Price}, Larry R.},
        title = "{Rapid Bayesian position reconstruction for gravitational-wave transients}",
      journal = {Physical Review D},
         year = 2016,
        month = jan,
       volume = {93},
       number = {2},
          eid = {024013},
        pages = {024013},
          doi = {10.1103/PhysRevD.93.024013},
archivePrefix = {arXiv},
       eprint = {1508.03634},
 primaryClass = {gr-qc},
       adsurl = {https://ui.adsabs.harvard.edu/abs/2016PhRvD..93b4013S}
}

@article{Veske_2020,
   title={Neutrino emission upper limits with maximum likelihood estimators for joint astrophysical neutrino searches with large sky localizations},
   volume={2020},
   ISSN={1475-7516},
   url={http://dx.doi.org/10.1088/1475-7516/2020/05/016},
   DOI={10.1088/1475-7516/2020/05/016},
   number={05},
   journal={Journal of Cosmology and Astroparticle Physics},
   publisher={IOP Publishing},
   author={Veske, Doğa and Márka, Zsuzsa and Bartos, Imre and Márka, Szabolcs},
   year={2020},
   month=may, pages={016–016} }

@ARTICLE{2020ApJ...905L..13D,
       author = {{de Bruijn}, Oliver and {Bartos}, Imre and {Biermann}, Peter L. and {Tjus}, J. Becker},
        title = "{Recurrent Neutrino Emission from Supermassive Black Hole Mergers}",
      journal = {The Astrophysical Journal Letters},
         year = 2020,
        month = dec,
       volume = {905},
       number = {1},
          eid = {L13},
        pages = {L13},
          doi = {10.3847/2041-8213/abc950},
archivePrefix = {arXiv},
       eprint = {2006.11288},
 primaryClass = {astro-ph.HE},
       adsurl = {https://ui.adsabs.harvard.edu/abs/2020ApJ...905L..13D}
}

@ARTICLE{2020MNRAS.492..843G,
       author = {{Guetta}, Dafne and {Rahin}, Roi and {Bartos}, Imre and {Della Valle}, Massimo},
        title = "{Constraining the fraction of core-collapse supernovae harbouring choked jets with high-energy neutrinos}",
      journal = {\mnras},
         year = 2020,
        month = feb,
       volume = {492},
       number = {1},
        pages = {843-847},
          doi = {10.1093/mnras/stz3245},
archivePrefix = {arXiv},
       eprint = {1906.07399},
 primaryClass = {astro-ph.HE},
       adsurl = {https://ui.adsabs.harvard.edu/abs/2020MNRAS.492..843G}
}

@INPROCEEDINGS{2019ICRC...36..865D,
       author = {{De Wasseige}, G. and {Bartos}, I. and {de Vries}, K. and {O'Sullivan}, E.},
        title = "{Probing neutrino emission at GeV energies from compact binary mergers with IceCube}",
    booktitle = {36th International Cosmic Ray Conference (ICRC2019)},
         year = 2019,
       series = {International Cosmic Ray Conference},
       volume = {36},
        month = jul,
          eid = {865},
        pages = {865},
          doi = {10.22323/1.358.0865},
archivePrefix = {arXiv},
       eprint = {1908.08299},
 primaryClass = {astro-ph.HE},
       adsurl = {https://ui.adsabs.harvard.edu/abs/2019ICRC...36..865D}
}

@ARTICLE{2019ApJ...878...34F,
       author = {{Fang}, Ke and {Metzger}, Brian D. and {Murase}, Kohta and {Bartos}, Imre and {Kotera}, Kumiko},
        title = "{Multimessenger Implications of AT2018cow: High-energy Cosmic-Ray and Neutrino Emissions from Magnetar-powered Superluminous Transients}",
      journal = {The Astrophysical Journal},
         year = 2019,
        month = jun,
       volume = {878},
       number = {1},
          eid = {34},
        pages = {34},
          doi = {10.3847/1538-4357/ab1b72},
archivePrefix = {arXiv},
       eprint = {1812.11673},
 primaryClass = {astro-ph.HE},
       adsurl = {https://ui.adsabs.harvard.edu/abs/2019ApJ...878...34F}
}

@ARTICLE{2018PhRvD..98d3020K,
       author = {{Kimura}, Shigeo S. and {Murase}, Kohta and {Bartos}, Imre and {Ioka}, Kunihito and {Heng}, Ik Siong and {M{\'e}sz{\'a}ros}, Peter},
        title = "{Transejecta high-energy neutrino emission from binary neutron star mergers}",
      journal = {Physical Review D},
         year = 2018,
        month = aug,
       volume = {98},
       number = {4},
          eid = {043020},
        pages = {043020},
          doi = {10.1103/PhysRevD.98.043020},
archivePrefix = {arXiv},
       eprint = {1805.11613},
 primaryClass = {astro-ph.HE},
       adsurl = {https://ui.adsabs.harvard.edu/abs/2018PhRvD..98d3020K}
}

@ARTICLE{2023ApJ...959...96A,
       author = {{Abbasi}, R. and {Ackermann}, M. and {Adams}, J. and {Agarwalla}, S.~K. and {Aguilar}, J.~A. and {Ahlers}, M. and {Alameddine}, J.~M. and {Amin}, N.~M. and {Andeen}, K. and {Anton}, G. and {Arg{\"u}elles}, C. and {Ashida}, Y. and {Athanasiadou}, S. and {Axani}, S.~N. and {Bai}, X. and {Balagopal V.}, A. and {Baricevic}, M. and {Barwick}, S.~W. and {Basu}, V. and {Bay}, R. and {Beatty}, J.~J. and {Becker}, K. -H. and {Becker Tjus}, J. and {Beise}, J. and {Bellenghi}, C. and {BenZvi}, S. and {Berley}, D. and {Bernardini}, E. and {Besson}, D.~Z. and {Binder}, G. and {Bindig}, D. and {Blaufuss}, E. and {Blot}, S. and {Bontempo}, F. and {Book}, J.~Y. and {Boscolo Meneguolo}, C. and {B{\"o}ser}, S. and {Botner}, O. and {B{\"o}ttcher}, J. and {Bourbeau}, E. and {Braun}, J. and {Brinson}, B. and {Brostean-Kaiser}, J. and {Burley}, R.~T. and {Busse}, R.~S. and {Butterfield}, D. and {Campana}, M.~A. and {Carloni}, K. and {Carnie-Bronca}, E.~G. and {Chattopadhyay}, S. and {Chau}, N. and {Chen}, C. and {Chen}, Z. and {Chirkin}, D. and {Choi}, S. and {Clark}, B.~A. and {Classen}, L. and {Coleman}, A. and {Collin}, G.~H. and {Connolly}, A. and {Conrad}, J.~M. and {Coppin}, P. and {Correa}, P. and {Countryman}, S. and {Cowen}, D.~F. and {Dave}, P. and {De Clercq}, C. and {DeLaunay}, J.~J. and {Delgado L{\'o}pez}, D. and {Dembinski}, H. and {Deoskar}, K. and {Desai}, A. and {Desiati}, P. and {de Vries}, K.~D. and {de Wasseige}, G. and {DeYoung}, T. and {Diaz}, A. and {D{\'\i}az-V{\'e}lez}, J.~C. and {Dittmer}, M. and {Domi}, A. and {Dujmovic}, H. and {DuVernois}, M.~A. and {Ehrhardt}, T. and {Eller}, P. and {Engel}, R. and {Erpenbeck}, H. and {Evans}, J. and {Evenson}, P.~A. and {Fan}, K.~L. and {Fang}, K. and {Fazely}, A.~R. and {Fedynitch}, A. and {Feigl}, N. and {Fiedlschuster}, S. and {Finley}, C. and {Fischer}, L. and {Fox}, D. and {Franckowiak}, A. and {Friedman}, E. and {Fritz}, A. and {F{\"u}rst}, P. and {Gaisser}, T.~K. and {Gallagher}, J. and {Ganster}, E. and {Garcia}, A. and {Gerhardt}, L. and {Ghadimi}, A. and {Glaser}, C. and {Glauch}, T. and {Gl{\"u}senkamp}, T. and {Goehlke}, N. and {Gonzalez}, J.~G. and {Goswami}, S. and {Grant}, D. and {Gray}, S.~J. and {Griffin}, S. and {Griswold}, S. and {G{\"u}nther}, C. and {Gutjahr}, P. and {Haack}, C. and {Hallgren}, A. and {Halliday}, R. and {Halve}, L. and {Halzen}, F. and {Hamdaoui}, H. and {Ha Minh}, M. and {Hanson}, K. and {Hardin}, J. and {Harnisch}, A.~A. and {Hatch}, P. and {Haungs}, A. and {Helbing}, K. and {Hellrung}, J. and {Henningsen}, F. and {Heuermann}, L. and {Heyer}, N. and {Hickford}, S. and {Hidvegi}, A. and {Hill}, C. and {Hill}, G.~C. and {Hoffman}, K.~D. and {Hoshina}, K. and {Hou}, W. and {Huber}, T. and {Hultqvist}, K. and {H{\"u}nnefeld}, M. and {Hussain}, R. and {Hymon}, K. and {In}, S. and {Ishihara}, A. and {Jacquart}, M. and {Jansson}, M. and {Japaridze}, G.~S. and {Jayakumar}, K. and {Jeong}, M. and {Jin}, M. and {Jones}, B.~J.~P. and {Kang}, D. and {Kang}, W. and {Kang}, X. and {Kappes}, A. and {Kappesser}, D. and {Kardum}, L. and {Karg}, T. and {Karl}, M. and {Karle}, A. and {Katz}, U. and {Kauer}, M. and {Kelley}, J.~L. and {Khatee Zathul}, A. and {Kheirandish}, A. and {Kiryluk}, J. and {Klein}, S.~R. and {Kochocki}, A. and {Koirala}, R. and {Kolanoski}, H. and {Kontrimas}, T. and {K{\"o}pke}, L. and {Kopper}, C. and {Koskinen}, D.~J. and {Koundal}, P. and {Kovacevich}, M. and {Kowalski}, M. and {Kozynets}, T. and {Kruiswijk}, K. and {Krupczak}, E. and {Kumar}, A. and {Kun}, E. and {Kurahashi}, N. and {Lad}, N. and {Lagunas Gualda}, C. and {Lamoureux}, M. and {Larson}, M.~J. and {Lauber}, F. and {Lazar}, J.~P. and {Lee}, J.~W. and {Leonard DeHolton}, K. and {Leszczy{\'n}ska}, A. and {Lincetto}, M. and {Liu}, Q.~R. and {Liubarska}, M. and {Lohfink}, E. and {Love}, C. and {Lozano Mariscal}, C.~J. and {Lu}, L. and {Lucarelli}, F. and {Ludwig}, A. and {Luszczak}, W. and {Lyu}, Y. and {Madsen}, J. and {Mahn}, K.~B.~M. and {Makino}, Y. and {Mancina}, S. and {Marie Sainte}, W. and {Mari{\c{s}}}, I.~C. and {Marka}, S. and {Marka}, Z. and {Marsee}, M. and {Martinez-Soler}, I. and {Maruyama}, R. and {Mayhew}, F. and {McElroy}, T. and {McNally}, F. and {Mead}, J.~V. and {Meagher}, K. and {Mechbal}, S. and {Medina}, A. and {Meier}, M. and {Meighen-Berger}, S. and {Merckx}, Y. and {Merten}, L. and {Micallef}, J. and {Montaruli}, T. and {Moore}, R.~W. and {Morii}, Y. and {Morse}, R. and {Moulai}, M. and {Mukherjee}, T. and {Naab}, R. and {Nagai}, R. and {Nakos}, M. and {Naumann}, U. and {Necker}, J. and {Neumann}, M. and {Niederhausen}, H. and {Nisa}, M.~U. and {Noell}, A. and {Nowicki}, S.~C. and {Obertacke Pollmann}, A. and {O'Dell}, V. and {Oehler}, M. and {Oeyen}, B. and {Olivas}, A. and {Orsoe}, R. and {Osborn}, J. and {O'Sullivan}, E. and {Pandya}, H. and {Park}, N. and {Parker}, G.~K. and {Paudel}, E.~N. and {Paul}, L. and {P{\'e}rez de los Heros}, C. and {Peterson}, J. and {Philippen}, S. and {Pieper}, S. and {Pizzuto}, A. and {Plum}, M. and {Pont{\'e}n}, A. and {Popovych}, Y. and {Prado Rodriguez}, M. and {Pries}, B. and {Procter-Murphy}, R. and {Przybylski}, G.~T. and {Rack-Helleis}, J. and {Rawlins}, K. and {Rechav}, Z. and {Rehman}, A. and {Reichherzer}, P. and {Renzi}, G. and {Resconi}, E. and {Reusch}, S. and {Rhode}, W. and {Richman}, M. and {Riedel}, B. and {Roberts}, E.~J. and {Robertson}, S. and {Rodan}, S. and {Roellinghoff}, G. and {Rongen}, M. and {Rott}, C. and {Ruhe}, T. and {Ruohan}, L. and {Ryckbosch}, D. and {Athanasiadou}, S. and {Safa}, I. and {Saffer}, J. and {Salazar-Gallegos}, D. and {Sampathkumar}, P. and {Sanchez Herrera}, S.~E. and {Sandrock}, A. and {Santander}, M. and {Sarkar}, S. and {Sarkar}, S. and {Savelberg}, J. and {Savina}, P. and {Schaufel}, M. and {Schieler}, H. and {Schindler}, S. and {Schl{\"u}ter}, B. and {Schl{\"u}ter}, F. and {Schmidt}, T. and {Schneider}, J. and {Schr{\"o}der}, F.~G. and {Schumacher}, L. and {Schwefer}, G. and {Sclafani}, S. and {Seckel}, D. and {Seunarine}, S. and {Sharma}, A. and {Shefali}, S. and {Shimizu}, N. and {Silva}, M. and {Skrzypek}, B. and {Smithers}, B. and {Snihur}, R. and {Soedingrekso}, J. and {S{\o}gaard}, A. and {Soldin}, D. and {Sommani}, G. and {Spannfellner}, C. and {Spiczak}, G.~M. and {Spiering}, C. and {Stamatikos}, M. and {Stanev}, T. and {Stezelberger}, T. and {St{\"u}rwald}, T. and {Stuttard}, T. and {Sullivan}, G.~W. and {Taboada}, I. and {Ter-Antonyan}, S. and {Thompson}, W.~G. and {Thwaites}, J. and {Tilav}, S. and {Tollefson}, K. and {T{\"o}nnis}, C. and {Toscano}, S. and {Tosi}, D. and {Trettin}, A. and {Tung}, C.~F. and {Turcotte}, R. and {Twagirayezu}, J.~P. and {Ty}, B. and {Unland Elorrieta}, M.~A. and {Upadhyay}, A.~K. and {Upshaw}, K. and {Valtonen-Mattila}, N. and {Vandenbroucke}, J. and {van Eijndhoven}, N. and {Vannerom}, D. and {van Santen}, J. and {Vara}, J. and {Veitch-Michaelis}, J. and {Venugopal}, M. and {Verpoest}, S. and {Veske}, D. and {Walck}, C. and {Watson}, T.~B. and {Weaver}, C. and {Weigel}, P. and {Weindl}, A. and {Weldert}, J. and {Wendt}, C. and {Werthebach}, J. and {Weyrauch}, M. and {Whitehorn}, N. and {Wiebusch}, C.~H. and {Willey}, N. and {Williams}, D.~R. and {Wolf}, M. and {Wrede}, G. and {Xu}, X.~W. and {Yanez}, J.~P. and {Yildizci}, E. and {Yoshida}, S. and {Yu}, F. and {Yu}, S. and {Yuan}, T. and {Zhang}, Z. and {Zhelnin}, P. and {IceCube Collaboration}},
        title = "{A Search for IceCube Sub-TeV Neutrinos Correlated with Gravitational-wave Events Detected By LIGO/Virgo}",
      journal = {The Astrophysical Journal},
         year = 2023,
        month = dec,
       volume = {959},
       number = {2},
          eid = {96},
        pages = {96},
          doi = {10.3847/1538-4357/aceefc},
archivePrefix = {arXiv},
       eprint = {2303.15970},
 primaryClass = {astro-ph.HE},
       adsurl = {https://ui.adsabs.harvard.edu/abs/2023ApJ...959...96A}
}

@article{ATHAR_2006,
   title={INTRINSIC AND OSCILLATED ASTROPHYSICAL NEUTRINO FLAVOR RATIOS REVISITED},
   volume={21},
   ISSN={1793-6632},
   url={http://dx.doi.org/10.1142/S021773230602038X},
   DOI={10.1142/s021773230602038x},
   number={13},
   journal={Modern Physics Letters A},
   publisher={World Scientific Pub Co Pte Lt},
   author={Athar, H. and Kim, C. S. and Lee, Jake},
   year={2006},
   month=apr, pages={1049–1065} }

@article{Pakvasa_2008,
   title={Flavor ratios of astrophysical neutrinos: implications for precision measurements},
   volume={2008},
   ISSN={1029-8479},
   url={http://dx.doi.org/10.1088/1126-6708/2008/02/005},
   DOI={10.1088/1126-6708/2008/02/005},
   number={02},
   journal={Journal of High Energy Physics},
   publisher={Springer Science and Business Media LLC},
   author={Pakvasa, Sandip and Rodejohann, Werner and Weiler, Thomas J},
   year={2008},
   month=feb, pages={005–005} }

@ARTICLE{2014ARA&A..52...43B,
       author = {{Berger}, Edo},
        title = "{Short-Duration Gamma-Ray Bursts}",
      journal = {\araa},
         year = 2014,
        month = aug,
       volume = {52},
        pages = {43-105},
          doi = {10.1146/annurev-astro-081913-035926},
archivePrefix = {arXiv},
       eprint = {1311.2603},
 primaryClass = {astro-ph.HE},
       adsurl = {https://ui.adsabs.harvard.edu/abs/2014ARA&A..52...43B}
}

@article{PhysRevX.13.011048,
  title = {Population of Merging Compact Binaries Inferred Using Gravitational Waves through GWTC-3},
  author = {Abbott, R. and Abbott, T. D. and Acernese, F. and Ackley, K. and Adams, C. and Adhikari, N. and Adhikari, R. X. and Adya, V. B. and Affeldt, C. and Agarwal, D. and others},
  collaboration = {LIGO Scientific Collaboration, Virgo Collaboration, and KAGRA Collaboration},
  journal = {Physical Review X},
  volume = {13},
  issue = {1},
  pages = {011048},
  numpages = {75},
  year = {2023},
  month = {Mar},
  publisher = {American Physical Society},
  doi = {10.1103/PhysRevX.13.011048},
  url = {https://link.aps.org/doi/10.1103/PhysRevX.13.011048}
}

@ARTICLE{2023ApJ...944...80A,
       author = {{Abbasi}, R. and {Ackermann}, M. and {Adams}, J. and {Aggarwal}, N. and {Aguilar}, J.~A. and {Ahlers}, M. and {Ahrens}, M. and {Alameddine}, J.~M. and {Alves}, A.~A. and {Amin}, N.~M. and {Andeen}, K. and {Anderson}, T. and {Anton}, G. and {Arg{\"u}elles}, C. and {Asali}, Y. and {Ashida}, Y. and {Athanasiadou}, S. and {Axani}, S. and {Bai}, X. and {Balagopal V.}, A. and {Baricevic}, M. and {Bartos}, I. and {Barwick}, S.~W. and {Basu}, V. and {Bay}, R. and {Beatty}, J.~J. and {Becker}, K. -H. and {Becker Tjus}, J. and {Beise}, J. and {Bellenghi}, C. and {Benda}, S. and {BenZvi}, S. and {Berley}, D. and {Bernardini}, E. and {Besson}, D.~Z. and {Binder}, G. and {Bindig}, D. and {Blaufuss}, E. and {Blot}, S. and {Bontempo}, F. and {Book}, J.~Y. and {Borowka}, J. and {B{\"o}ser}, S. and {Botner}, O. and {B{\"o}ttcher}, J. and {Bourbeau}, E. and {Bradascio}, F. and {Braun}, J. and {Brinson}, B. and {Bron}, S. and {Brostean-Kaiser}, J. and {Burley}, R.~T. and {Busse}, R.~S. and {Campana}, M.~A. and {Carnie-Bronca}, E.~G. and {Chen}, C. and {Chen}, Z. and {Chirkin}, D. and {Choi}, K. and {Clark}, B.~A. and {Classen}, L. and {Coleman}, A. and {Collin}, G.~H. and {Connolly}, A. and {Conrad}, J.~M. and {Coppin}, P. and {Correa}, P. and {Countryman}, S.~T. and {Cowen}, D.~F. and {Cross}, R. and {Dappen}, C. and {Dave}, P. and {De Clercq}, C. and {DeLaunay}, J.~J. and {Delgado L{\'o}pez}, D. and {Dembinski}, H. and {Deoskar}, K. and {Desai}, A. and {Desiati}, P. and {de Vries}, K.~D. and {de Wasseige}, G. and {DeYoung}, T. and {Diaz}, A. and {D{\'\i}az-V{\'e}lez}, J.~C. and {Dittmer}, M. and {Dujmovic}, H. and {DuVernois}, M.~A. and {Ehrhardt}, T. and {Eller}, P. and {Engel}, R. and {Erpenbeck}, H. and {Evans}, J. and {Evenson}, P.~A. and {Fan}, K.~L. and {Fazely}, A.~R. and {Fedynitch}, A. and {Feigl}, N. and {Fiedlschuster}, S. and {Fienberg}, A.~T. and {Finley}, C. and {Fischer}, L. and {Fox}, D. and {Franckowiak}, A. and {Friedman}, E. and {Fritz}, A. and {F{\"u}rst}, P. and {Gaisser}, T.~K. and {Gallagher}, J. and {Ganster}, E. and {Garcia}, A. and {Garrappa}, S. and {Gerhardt}, L. and {Ghadimi}, A. and {Glaser}, C. and {Glauch}, T. and {Gl{\"u}senkamp}, T. and {Goehlke}, N. and {Gonzalez}, J.~G. and {Goswami}, S. and {Grant}, D. and {Gr{\'e}goire}, T. and {Griswold}, S. and {G{\"u}nther}, C. and {Gutjahr}, P. and {Haack}, C. and {Hallgren}, A. and {Halliday}, R. and {Halve}, L. and {Halzen}, F. and {Hamdaoui}, H. and {Ha Minh}, M. and {Hanson}, K. and {Hardin}, J. and {Harnisch}, A.~A. and {Hatch}, P. and {Haungs}, A. and {Helbing}, K. and {Hellrung}, J. and {Henningsen}, F. and {Heuermann}, L. and {Hickford}, S. and {Hill}, C. and {Hill}, G.~C. and {Hoffman}, K.~D. and {Hoshina}, K. and {Hou}, W. and {Huber}, T. and {Hultqvist}, K. and {H{\"u}nnefeld}, M. and {Hussain}, R. and {Hymon}, K. and {In}, S. and {Iovine}, N. and {Ishihara}, A. and {Jansson}, M. and {Japaridze}, G.~S. and {Jeong}, M. and {Jin}, M. and {Jones}, B.~J.~P. and {Kang}, D. and {Kang}, W. and {Kang}, X. and {Kappes}, A. and {Kappesser}, D. and {Kardum}, L. and {Karg}, T. and {Karl}, M. and {Karle}, A. and {Katz}, U. and {Kauer}, M. and {Kelley}, J.~L. and {Kheirandish}, A. and {Kin}, K. and {Kiryluk}, J. and {Klein}, S.~R. and {Kochocki}, A. and {Koirala}, R. and {Kolanoski}, H. and {Kontrimas}, T. and {K{\"o}pke}, L. and {Kopper}, C. and {Koskinen}, D.~J. and {Koundal}, P. and {Kovacevich}, M. and {Kowalski}, M. and {Kozynets}, T. and {Krupczak}, E. and {Kun}, E. and {Kurahashi}, N. and {Lad}, N. and {Lagunas Gualda}, C. and {Larson}, M.~J. and {Lauber}, F. and {Lazar}, J.~P. and {Lee}, J.~W. and {Leonard}, K. and {Leszczy{\'n}ska}, A. and {Lincetto}, M. and {Liu}, Q.~R. and {Liubarska}, M. and {Lohfink}, E. and {Love}, C. and {Mariscal}, C.~J. Lozano and {Lu}, L. and {Lucarelli}, F. and {Ludwig}, A. and {Luszczak}, W. and {Lyu}, Y. and {Ma}, W.~Y. and {Madsen}, J. and {Mahn}, K.~B.~M. and {Makino}, Y. and {Mancina}, S. and {Sainte}, W. Marie and {Mari{\c{s}}}, I.~C. and {M{\'a}rka}, S. and {M{\'a}rka}, Z. and {Marsee}, M. and {Martinez-Soler}, I. and {Maruyama}, R. and {McElroy}, T. and {McNally}, F. and {Mead}, J.~V. and {Meagher}, K. and {Mechbal}, S. and {Medina}, A. and {Meier}, M. and {Meighen-Berger}, S. and {Merckx}, Y. and {Micallef}, J. and {Mockler}, D. and {Montaruli}, T. and {Moore}, R.~W. and {Morse}, R. and {Moulai}, M. and {Mukherjee}, T. and {Naab}, R. and {Nagai}, R. and {Naumann}, U. and {Necker}, J. and {Neumann}, M. and {Niederhausen}, H. and {Nisa}, M.~U. and {Nowicki}, S.~C. and {Obertacke Pollmann}, A. and {Oehler}, M. and {Oeyen}, B. and {Olivas}, A. and {Orsoe}, R. and {Osborn}, J. and {O'Sullivan}, E. and {Pandya}, H. and {Pankova}, D.~V. and {Park}, N. and {Parker}, G.~K. and {Paudel}, E.~N. and {Paul}, L. and {P{\'e}rez de los Heros}, C. and {Peters}, L. and {Peterson}, J. and {Philippen}, S. and {Pieper}, S. and {Pizzuto}, A. and {Plum}, M. and {Popovych}, Y. and {Porcelli}, A. and {Prado Rodriguez}, M. and {Pries}, B. and {Przybylski}, G.~T. and {Raab}, C. and {Rack-Helleis}, J. and {Rameez}, M. and {Rawlins}, K. and {Rechav}, Z. and {Rehman}, A. and {Reichherzer}, P. and {Renzi}, G. and {Resconi}, E. and {Reusch}, S. and {Rhode}, W. and {Richman}, M. and {Riedel}, B. and {Roberts}, E.~J. and {Robertson}, S. and {Rodan}, S. and {Roellinghoff}, G. and {Rongen}, M. and {Rott}, C. and {Ruhe}, T. and {Ruohan}, L. and {Ryckbosch}, D. and {Rysewyk Cantu}, D. and {Safa}, I. and {Saffer}, J. and {Salazar-Gallegos}, D. and {Sampathkumar}, P. and {Sanchez Herrera}, S.~E. and {Sandrock}, A. and {Santander}, M. and {Sarkar}, S. and {Sarkar}, S. and {Satalecka}, K. and {Schaufel}, M. and {Schieler}, H. and {Schindler}, S. and {Schlueter}, B. and {Schmidt}, T. and {Schneider}, J. and {Schr{\"o}der}, F.~G. and {Schumacher}, L. and {Schwefer}, G. and {Sclafani}, S. and {Seckel}, D. and {Seunarine}, S. and {Sharma}, A. and {Shefali}, S. and {Shimizu}, N. and {Silva}, M. and {Silva Oliveira}, A.~C. and {Skrzypek}, B. and {Smithers}, B. and {Snihur}, R. and {Soedingrekso}, J. and {Sogaard}, A. and {Soldin}, D. and {Spannfellner}, C. and {Spiczak}, G.~M. and {Spiering}, C. and {Stamatikos}, M. and {Stanev}, T. and {Stein}, R. and {Stezelberger}, T. and {St{\"u}rwald}, T. and {Stuttard}, T. and {Sullivan}, A.~G. and {Sullivan}, G.~W. and {Taboada}, I. and {Ter-Antonyan}, S. and {Thompson}, W.~G. and {Thwaites}, J. and {Tilav}, S. and {Tollefson}, K. and {T{\"o}nnis}, C. and {Toscano}, S. and {Tosi}, D. and {Trettin}, A. and {Tung}, C.~F. and {Turcotte}, R. and {Twagirayezu}, J.~P. and {Ty}, B. and {Unland Elorrieta}, M.~A. and {Upshaw}, K. and {Valtonen-Mattila}, N. and {Vandenbroucke}, J. and {van Eijndhoven}, N. and {Vannerom}, D. and {van Santen}, J. and {Vara}, J. and {Veitch-Michaelis}, J. and {Verpoest}, S. and {Veske}, D. and {Walck}, C. and {Wang}, W. and {Watson}, T.~B. and {Weaver}, C. and {Weigel}, P. and {Weindl}, A. and {Weldert}, J. and {Wendt}, C. and {Werthebach}, J. and {Weyrauch}, M. and {Whitehorn}, N. and {Wiebusch}, C.~H. and {Willey}, N. and {Williams}, D.~R. and {Wolf}, M. and {Wrede}, G. and {Wulff}, J. and {Xu}, X.~W. and {Yanez}, J.~P. and {Yildizci}, E. and {Yoshida}, S. and {Yu}, S. and {Yuan}, T. and {Zhang}, Z. and {Zhelnin}, P. and {IceCube Collaboration}},
        title = "{IceCube Search for Neutrinos Coincident with Gravitational Wave Events from LIGO/Virgo Run O3}",
      journal = {The Astrophysical Journal},
         year = 2023,
        month = feb,
       volume = {944},
       number = {1},
          eid = {80},
        pages = {80},
          doi = {10.3847/1538-4357/aca5fc},
archivePrefix = {arXiv},
       eprint = {2208.09532},
 primaryClass = {astro-ph.HE},
       adsurl = {https://ui.adsabs.harvard.edu/abs/2023ApJ...944...80A}
}

@INPROCEEDINGS{2022APS..APRK14005M,
       author = {{Marka}, Zsuzsanna and {Oliveira}, Ana Carolina and {Veske}, Doga and {Hussain}, Raamis and {Balagopal V.}, Aswathi and {Vandenbroucke}, Justin},
        title = "{IceCube search for neutrinos coincident with gravitational wave detections reported for LIGO-Virgo's third observing run}",
    booktitle = {APS April Meeting Abstracts},
         year = 2022,
       series = {APS Meeting Abstracts},
       volume = {2022},
        month = apr,
          eid = {K14.005},
        pages = {K14.005},
       adsurl = {https://ui.adsabs.harvard.edu/abs/2022APS..APRK14005M}
}

@INPROCEEDINGS{2022icrc.confE.950V,
       author = {{Veske}, D. and {IceCube} and {Abbasi}, R. and {Ackermann}, M. and {Adams}, J. and {Aguilar}, J. and {Ahlers}, M. and {Ahrens}, M. and {Alispach}, C.~M. and {Alves Junior}, A.~A. and {Amin}, N.~M.~B. and {An}, R. and {Andeen}, K. and {Anderson}, T. and {Anton}, G. and {Arguelles}, C. and {Ashida}, Y. and {Axani}, S. and {Bai}, X. and {Balagopal V.}, A. and {Barbano}, A.~M. and {Barwick}, S.~W. and {Bastian}, B. and {Basu}, V. and {Baur}, S. and {Bay}, R.~C. and {Beatty}, J.~J. and {Becker}, K.~H. and {Becker Tjus}, J. and {Bellenghi}, C. and {BenZvi}, S. and {Berley}, D. and {Bernardini}, E. and {Besson}, D.~Z. and {Binder}, G. and {Bindig}, D. and {Blaufuss}, E. and {Blot}, S. and {Boddenberg}, M. and {Bontempo}, F. and {Borowka}, J. and {Boser}, S. and {Botner}, O. and {Bottcher}, J. and {Bourbeau}, E. and {Bradascio}, F. and {Braun}, J. and {Bron}, S. and {Brostean-Kaiser}, J. and {Browne}, S.~A. and {Burgman}, A. and {Burley}, R. and {Busse}, R. and {Campana}, M. and {Carnie-Bronca}, E. and {Chen}, C. and {Chirkin}, D. and {Choi}, K. and {Clark}, B. and {Clark}, K. and {Classen}, L. and {Coleman}, A. and {Collin}, G. and {Conrad}, J.~M. and {Coppin}, P. and {Correa}, P. and {Cowen}, D.~F. and {Cross}, R. and {Dappen}, C. and {Dave}, P. and {De Clercq}, C. and {DeLaunay}, J. and {Dembinski}, H. and {Deoskar}, K. and {De Ridder}, S. and {Desai}, A. and {Desiati}, P. and {de Vries}, K. and {de Wasseige}, G. and {De With}, M. and {DeYoung}, T. and {Dharani}, S. and {Diaz}, A. and {Diaz-Velez}, J.~C. and {Dittmer}, M. and {Dujmovic}, H. and {Dunkman}, M. and {DuVernois}, M. and {Dvorak}, E. and {Ehrhardt}, T. and {Eller}, P. and {Engel}, R. and {Erpenbeck}, H. and {Evans}, J. and {Evenson}, P.~A. and {Fan}, K.~L. and {Fazely}, A.~R. and {Fiedlschuster}, S. and {Fienberg}, A. and {Filimonov}, K. and {Finley}, C. and {Fischer}, L. and {Fox}, D.~B. and {Franckowiak}, A. and {Friedman}, E. and {Fritz}, A. and {Furst}, P. and {Gaisser}, T.~K. and {Gallagher}, J. and {Ganster}, E. and {Garcia}, A. and {Garrappa}, S. and {Gerhardt}, L. and {Ghadimi}, A. and {Glaser}, C. and {Glauch}, T. and {Glusenkamp}, T. and {Goldschmidt}, A. and {Gonzalez}, J. and {Goswami}, S. and {Grant}, D. and {Gr{\'e}goire}, T. and {Griswold}, S. and {Gunduz}, M. and {G{\"u}nther}, C. and {Haack}, C. and {Hallgren}, A. and {Halliday}, R. and {Halve}, L. and {Halzen}, F. and {Minh}, M. Ha and {Hanson}, K. and {Hardin}, J. and {Harnisch}, A.~A. and {Haungs}, A. and {Hauser}, S. and {Hebecker}, D. and {Helbing}, K. and {Henningsen}, F. and {Hettinger}, E.~C. and {Hickford}, S. and {Hignight}, J. and {Hill}, C. and {Hill}, G.~C. and {Hoffman}, K. and {Hoffmann}, R. and {Hoinka}, T. and {Hokanson-Fasig}, B. and {Hoshina}, K. and {Huang}, F. and {Huber}, M. and {Huber}, T. and {Hultqvist}, K. and {Hunnefeld}, M. and {Hussain}, R. and {In}, S. and {Iovine}, N. and {Ishihara}, A. and {Jansson}, M. and {Japaridze}, G. and {Jeong}, M. and {Jones}, B. and {Kang}, D. and {Kang}, W. and {Kang}, X. and {Kappes}, A. and {Kappesser}, D. and {Karg}, T. and {Karl}, M. and {Karle}, A. and {Katz}, U. and {Kauer}, M. and {Kellermann}, M. and {Kelley}, J.~L. and {Kheirandish}, A. and {Kin}, K. i. and {Kintscher}, T. and {Kiryluk}, J. and {Klein}, S. and {Koirala}, R. and {Kolanoski}, H. and {Kontrimas}, T. and {Kopke}, L. and {Kopper}, C. and {Kopper}, S. and {Koskinen}, D.~J. and {Koundal}, P. and {Kovacevich}, M. and {Kowalski}, M. and {Kozynets}, T. and {Kun}, E. and {Kurahashi}, N. and {Lad}, N. and {Lagunas Gualda}, C. and {Lanfranchi}, J. and {Larson}, M.~J. and {Lauber}, F.~H. and {Lazar}, J. and {Lee}, J. and {Leonard}, K. and {Leszczy{\'n}ska}, A. and {Li}, Y. and {Lincetto}, M. and {Liu}, Q. and {Liubarska}, M. and {Lohfink}, E. and {Lozano Mariscal}, C.~J. and {Lu}, L. and {Lucarelli}, F. and {Ludwig}, A. and {Luszczak}, W. and {Lyu}, Y. and {Ma}, W.~Y. and {Madsen}, J. and {Mahn}, K. and {Makino}, Y. and {Mancina}, S. and {Maris}, I.~C. and {Maruyama}, R.~H. and {Mase}, K. and {McElroy}, T. and {McNally}, F. and {Mead}, J.~V. and {Meagher}, K. and {Medina}, A. and {Meier}, M. and {Meighen-Berger}, S. and {Micallef}, J. and {Mockler}, D. and {Montaruli}, T. and {Moore}, R. and {Morse}, R. and {Moulai}, M. and {Naab}, R. and {Nagai}, R. and {Naumann}, U. and {Necker}, J. and {Nguyen}, L.~V. and {Niederhausen}, H. and {Nisa}, M. and {Nowicki}, S. and {Nygren}, D. and {Obertacke Pollmann}, A. and {Oehler}, M. and {Olivas}, A. and {O'Sullivan}, E. and {Pandya}, H. and {Pankova}, D. and {Park}, N. and {Parker}, G. and {Paudel}, E.~N. and {Paul}, L. and {Perez de los Heros}, C. and {Peters}, L. and {Peterson}, J. and {Philippen}, S. and {Pieloth}, D. and {Pieper}, S. and {Pittermann}, M. and {Pizzuto}, A. and {Plum}, M. and {Popovych}, Y. and {Porcelli}, A. and {Prado Rodriguez}, M. and {Price}, P.~B. and {Pries}, B. and {Przybylski}, G. and {Raab}, C. and {Raissi}, A. and {Rameez}, M. and {Rawlins}, K. and {Rea}, I.~C. and {Rehman}, A. and {Reichherzer}, P. and {Reimann}, R. and {Renzi}, G. and {Resconi}, E. and {Reusch}, S. and {Rhode}, W. and {Richman}, M. and {Riedel}, B. and {Roberts}, E. and {Robertson}, S. and {Roellinghoff}, G. and {Rongen}, M. and {Rott}, C. and {Ruhe}, T. and {Ryckbosch}, D. and {Rysewyk Cantu}, D. and {Safa}, I. and {Saffer}, J. and {Sanchez Herrera}, S. and {Sandrock}, A. and {Sandroos}, J. and {Santander}, M. and {Sarkar}, S. and {Sarkar}, S. and {Satalecka}, K. and {Scharf}, M.~K. and {Schaufel}, M. and {Schieler}, H. and {Schindler}, S. and {Schlunder}, P. and {Schmidt}, T. and {Schneider}, A. and {Schneider}, J. and {Schr{\"o}der}, F.~G. and {Schumacher}, L.~J. and {Schwefer}, G. and {Sclafani}, S. and {Seckel}, D. and {Seunarine}, S. and {Sharma}, A. and {Shefali}, S. and {Silva}, M. and {Skrzypek}, B. and {Smithers}, B. and {Snihur}, R. and {Soedingrekso}, J. and {Soldin}, D. and {Spannfellner}, C. and {Spiczak}, G. and {Spiering}, C. and {Stachurska}, J. and {Stamatikos}, M. and {Stanev}, T. and {Stein}, R. and {Stettner}, J. and {Steuer}, A. and {Stezelberger}, T. and {Sturwald}, T. and {Stuttard}, T. and {Sullivan}, G.~W. and {Taboada}, I. and {Tenholt}, F. and {Ter-Antonyan}, S. and {Tilav}, S. and {Tischbein}, F. and {Tollefson}, K. and {Tomankova}, L. and {T{\"o}nnis}, C. and {Toscano}, S. and {Tosi}, D. and {Trettin}, A. and {Tselengidou}, M. and {Tung}, C. and {Turcati}, A. and {Turcotte}, R. and {Turley}, C. and {Twagirayezu}, J.~P. and {Ty}, B. and {Unland Elorrieta}, M. and {Valtonen-Mattila}, N. and {Vandenbroucke}, J. and {van Eijndhoven}, N. and {Vannerom}, D. and {van Santen}, J. and {Verpoest}, S. and {Vraeghe}, M. and {Walck}, C. and {Watson}, T. and {Weaver}, C. and {Weigel}, P. and {Weindl}, A. and {Weiss}, M. and {Weldert}, J. and {Wendt}, C. and {Werthebach}, J. and {Weyrauch}, M. and {Whitehorn}, N. and {Wiebusch}, C.~H. and {Williams}, D. and {Wolf}, M. and {Woschnagg}, K. and {Wrede}, G. and {Wulff}, J. and {Xu}, X. and {Xu}, Y. and {Yanez}, J.~P. and {Yoshida}, S. and {Yu}, S. and {Yuan}, T. and {Zhang}, Z. and {Marka}, Z. and {Countryman}, S. and {Asali}, Y. and {Silva Oliveira}, A.},
        title = "{Multi-messenger searches via IceCube's high-energy neutrinos and gravitational-wave detections of LIGO/Virgo}",
    booktitle = {37th International Cosmic Ray Conference},
         year = 2022,
        month = mar,
          eid = {950},
        pages = {950},
          doi = {10.22323/1.395.0950},
archivePrefix = {arXiv},
       eprint = {2107.09663},
 primaryClass = {astro-ph.HE},
       adsurl = {https://ui.adsabs.harvard.edu/abs/2022icrc.confE.950V}
}

@ARTICLE{2021ApJ...909..126K,
       author = {{Keivani}, Azadeh and {Kennea}, Jamie A. and {Evans}, Phil A. and {Tohuvavohu}, Aaron and {Rapisura}, Riki and {Oates}, Samantha R. and {Countryman}, Stefan and {Bartos}, Imre and {M{\'a}rka}, Zsuzsa and {Veske}, Do{\u{g}}a and {M{\'a}rka}, Szabolcs and {Fox}, Derek B.},
        title = "{Swift Follow-up Observations of Gravitational-wave and High-energy Neutrino Coincident Signals}",
      journal = {The Astrophysical Journal},
         year = 2021,
        month = mar,
       volume = {909},
       number = {2},
          eid = {126},
        pages = {126},
          doi = {10.3847/1538-4357/abdab4},
archivePrefix = {arXiv},
       eprint = {2011.01319},
 primaryClass = {astro-ph.HE},
       adsurl = {https://ui.adsabs.harvard.edu/abs/2021ApJ...909..126K}
}

@ARTICLE{2020ApJ...898L..10A,
       author = {{Aartsen}, M.~G. and {Ackermann}, M. and {Adams}, J. and {Aguilar}, J.~A. and {Ahlers}, M. and {Ahrens}, M. and {Alispach}, C. and {Andeen}, K. and {Anderson}, T. and {Ansseau}, I. and {Anton}, G. and {Arg{\"u}elles}, C. and {Auffenberg}, J. and {Axani}, S. and {Bagherpour}, H. and {Bai}, X. and {A. Balagopal}, V. and {Barbano}, A. and {Bartos}, I. and {Barwick}, S.~W. and {Bastian}, B. and {Baum}, V. and {Baur}, S. and {Bay}, R. and {Beatty}, J.~J. and {Becker}, K. -H. and {Tjus}, J. Becker and {BenZvi}, S. and {Berley}, D. and {Bernardini}, E. and {Besson}, D.~Z. and {Binder}, G. and {Bindig}, D. and {Blaufuss}, E. and {Blot}, S. and {Bohm}, C. and {B{\"o}ser}, S. and {Botner}, O. and {B{\"o}ttcher}, J. and {Bourbeau}, E. and {Bourbeau}, J. and {Bradascio}, F. and {Braun}, J. and {Bron}, S. and {Brostean-Kaiser}, J. and {Burgman}, A. and {Buscher}, J. and {Busse}, R.~S. and {Carver}, T. and {Chen}, C. and {Cheung}, E. and {Chirkin}, D. and {Choi}, S. and {Clark}, B.~A. and {Clark}, K. and {Classen}, L. and {Coleman}, A. and {Collin}, G.~H. and {Conrad}, J.~M. and {Coppin}, P. and {Corley}, K.~R. and {Correa}, P. and {Countryman}, S. and {Cowen}, D.~F. and {Cross}, R. and {Dave}, P. and {Clercq}, C. De and {DeLaunay}, J.~J. and {Dembinski}, H. and {Deoskar}, K. and {Ridder}, S. De and {Desiati}, P. and {Vries}, K.~D. de and {Wasseige}, G. de and {With}, M. de and {DeYoung}, T. and {Diaz}, A. and {D{\'\i}az-V{\'e}lez}, J.~C. and {Dujmovic}, H. and {Dunkman}, M. and {Dvorak}, E. and {Eberhardt}, B. and {Ehrhardt}, T. and {Eller}, P. and {Engel}, R. and {Evenson}, P.~A. and {Fahey}, S. and {Fazely}, A.~R. and {Felde}, J. and {Filimonov}, K. and {Finley}, C. and {Fox}, D. and {Franckowiak}, A. and {Friedman}, E. and {Fritz}, A. and {Gaisser}, T.~K. and {Gallagher}, J. and {Ganster}, E. and {Garrappa}, S. and {Gerhardt}, L. and {Ghorbani}, K. and {Glauch}, T. and {Gl{\"u}senkamp}, T. and {Goldschmidt}, A. and {Gonzalez}, J.~G. and {Grant}, D. and {Gr{\'e}goire}, T. and {Griffith}, Z. and {Griswold}, S. and {G{\"u}nder}, M. and {G{\"u}nd{\"u}z}, M. and {Haack}, C. and {Hallgren}, A. and {Halliday}, R. and {Halve}, L. and {Halzen}, F. and {Hanson}, K. and {Haungs}, A. and {Hebecker}, D. and {Heereman}, D. and {Heix}, P. and {Helbing}, K. and {Hellauer}, R. and {Henningsen}, F. and {Hickford}, S. and {Hignight}, J. and {Hill}, G.~C. and {Hoffman}, K.~D. and {Hoffmann}, R. and {Hoinka}, T. and {Hokanson-Fasig}, B. and {Hoshina}, K. and {Huang}, F. and {Huber}, M. and {Huber}, T. and {Hultqvist}, K. and {H{\"u}nnefeld}, M. and {Hussain}, R. and {In}, S. and {Iovine}, N. and {Ishihara}, A. and {Jansson}, M. and {Japaridze}, G.~S. and {Jeong}, M. and {Jero}, K. and {Jones}, B.~J.~P. and {Jonske}, F. and {Joppe}, R. and {Kang}, D. and {Kang}, W. and {Kappes}, A. and {Kappesser}, D. and {Karg}, T. and {Karl}, M. and {Karle}, A. and {Katz}, U. and {Kauer}, M. and {Keivani}, A. and {Kellermann}, M. and {Kelley}, J.~L. and {Kheirandish}, A. and {Kim}, J. and {Kintscher}, T. and {Kiryluk}, J. and {Kittler}, T. and {Klein}, S.~R. and {Koirala}, R. and {Kolanoski}, H. and {K{\"o}pke}, L. and {Kopper}, C. and {Kopper}, S. and {Koskinen}, D.~J. and {Kowalski}, M. and {Krings}, K. and {Kr{\"u}ckl}, G. and {Kulacz}, N. and {Kurahashi}, N. and {Kyriacou}, A. and {Lanfranchi}, J.~L. and {Larson}, M.~J. and {Lauber}, F. and {Lazar}, J.~P. and {Leonard}, K. and {Leszczy{\'n}ska}, A. and {Liu}, Q.~R. and {Lohfink}, E. and {Mariscal}, C.~J. Lozano and {Lu}, L. and {Lucarelli}, F. and {Ludwig}, A. and {L{\"u}nemann}, J. and {Luszczak}, W. and {Lyu}, Y. and {Ma}, W.~Y. and {Madsen}, J. and {Maggi}, G. and {Mahn}, K.~B.~M. and {Makino}, Y. and {Mallik}, P. and {Mallot}, K. and {Mancina}, S. and {Mari{\c{s}}}, I.~C. and {Marka}, S. and {Marka}, Z. and {Maruyama}, R. and {Mase}, K. and {Maunu}, R. and {McNally}, F. and {Meagher}, K. and {Medici}, M. and {Medina}, A. and {Meier}, M. and {Meighen-Berger}, S. and {Merino}, G. and {Meures}, T. and {Micallef}, J. and {Mockler}, D. and {Moment{\'e}}, G. and {Montaruli}, T. and {Moore}, R.~W. and {Morse}, R. and {Moulai}, M. and {Muth}, P. and {Nagai}, R. and {Naumann}, U. and {Neer}, G. and {Nguyen}, L.~V. and {Niederhausen}, H. and {Nisa}, M.~U. and {Nowicki}, S.~C. and {Nygren}, D.~R. and {Pollmann}, A. Obertacke and {Oehler}, M. and {Olivas}, A. and {O'Murchadha}, A. and {O'Sullivan}, E. and {Palczewski}, T. and {Pandya}, H. and {Pankova}, D.~V. and {Park}, N. and {Peiffer}, P. and {de los Heros}, C. P{\'e}rez and {Philippen}, S. and {Pieloth}, D. and {Pieper}, S. and {Pinat}, E. and {Pizzuto}, A. and {Plum}, M. and {Porcelli}, A. and {Price}, P.~B. and {Przybylski}, G.~T. and {Raab}, C. and {Raissi}, A. and {Rameez}, M. and {Rauch}, L. and {Rawlins}, K. and {Rea}, I.~C. and {Rehman}, A. and {Reimann}, R. and {Relethford}, B. and {Renschler}, M. and {Renzi}, G. and {Resconi}, E. and {Rhode}, W. and {Richman}, M. and {Robertson}, S. and {Rongen}, M. and {Rott}, C. and {Ruhe}, T. and {Ryckbosch}, D. and {Cantu}, D. Rysewyk and {Safa}, I. and {Herrera}, S.~E. Sanchez and {Sandrock}, A. and {Sandroos}, J. and {Santander}, M. and {Sarkar}, S. and {Sarkar}, S. and {Satalecka}, K. and {Schaufel}, M. and {Schieler}, H. and {Schlunder}, P. and {Schmidt}, T. and {Schneider}, A. and {Schneider}, J. and {Schr{\"o}der}, F.~G. and {Schumacher}, L. and {Sclafani}, S. and {Seckel}, D. and {Seunarine}, S. and {Shefali}, S. and {Silva}, M. and {Snihur}, R. and {Soedingrekso}, J. and {Soldin}, D. and {Song}, M. and {Spiczak}, G.~M. and {Spiering}, C. and {Stachurska}, J. and {Stamatikos}, M. and {Stanev}, T. and {Stein}, R. and {Stettner}, J. and {Steuer}, A. and {Stezelberger}, T. and {Stokstad}, R.~G. and {St{\"o}{\ss}l}, A. and {Strotjohann}, N.~L. and {St{\"u}rwald}, T. and {Stuttard}, T. and {Sullivan}, G.~W. and {Taboada}, I. and {Tenholt}, F. and {Ter-Antonyan}, S. and {Terliuk}, A. and {Tilav}, S. and {Tollefson}, K. and {Tomankova}, L. and {T{\"o}nnis}, C. and {Toscano}, S. and {Tosi}, D. and {Trettin}, A. and {Tselengidou}, M. and {Tung}, C.~F. and {Turcati}, A. and {Turcotte}, R. and {Turley}, C.~F. and {Ty}, B. and {Unger}, E. and {Elorrieta}, M.~A. Unland and {Usner}, M. and {Vandenbroucke}, J. and {Driessche}, W. Van and {Eijk}, D. van and {Eijndhoven}, N. van and {Santen}, J. van and {Verpoest}, S. and {Veske}, D. and {Vraeghe}, M. and {Walck}, C. and {Wallace}, A. and {Wallraff}, M. and {Wandkowsky}, N. and {Watson}, T.~B. and {Weaver}, C. and {Weindl}, A. and {Weiss}, M.~J. and {Weldert}, J. and {Wendt}, C. and {Werthebach}, J. and {Whelan}, B.~J. and {Whitehorn}, N. and {Wiebe}, K. and {Wiebusch}, C.~H. and {Wille}, L. and {Williams}, D.~R. and {Wills}, L. and {Wolf}, M. and {Wood}, J. and {Wood}, T.~R. and {Woschnagg}, K. and {Wrede}, G. and {Xu}, D.~L. and {Xu}, X.~W. and {Xu}, Y. and {Yanez}, J.~P. and {Yodh}, G. and {Yoshida}, S. and {Yuan}, T. and {Z{\"o}cklein}, M.},
        title = "{IceCube Search for Neutrinos Coincident with Compact Binary Mergers from LIGO-Virgo's First Gravitational-wave Transient Catalog}",
      journal = {The Astrophysical Journal Letters},
         year = 2020,
        month = jul,
       volume = {898},
       number = {1},
          eid = {L10},
        pages = {L10},
          doi = {10.3847/2041-8213/ab9d24},
archivePrefix = {arXiv},
       eprint = {2004.02910},
 primaryClass = {astro-ph.HE},
       adsurl = {https://ui.adsabs.harvard.edu/abs/2020ApJ...898L..10A}
}

@INPROCEEDINGS{2019ICRC...36..930K,
       author = {{Keivani}, A. and {Veske}, D. and {Countryman}, S. and {Bartos}, I. and {Corely}, K.~R. and {Marka}, Z. and {Marka}, S.},
        title = "{Multi-messenger Gravitational-Wave + High-Energy Neutrino Searches with LIGO, Virgo and IceCube}",
    booktitle = {36th International Cosmic Ray Conference (ICRC2019)},
         year = 2019,
       series = {International Cosmic Ray Conference},
       volume = {36},
        month = jul,
          eid = {930},
        pages = {930},
          doi = {10.22323/1.358.0930},
archivePrefix = {arXiv},
       eprint = {1908.04996},
 primaryClass = {astro-ph.HE},
       adsurl = {https://ui.adsabs.harvard.edu/abs/2019ICRC...36..930K}
}

@article{galacticplane,
   title={Observation of high-energy neutrinos from the Galactic plane},
   volume={380},
   ISSN={1095-9203},
   url={http://dx.doi.org/10.1126/science.adc9818},
   DOI={10.1126/science.adc9818},
   number={6652},
   journal={Science},
   publisher={American Association for the Advancement of Science (AAAS)},
   author={Abbasi, R. and Ackermann, M. and Adams, J. and Aguilar, J. A. and Ahlers, M. and Ahrens, M. and Alameddine, J. M. and Alves, A. A. and Amin, N. M. and Andeen, K. and Anderson, T. and Anton, G. and Argüelles, C. and Ashida, Y. and Athanasiadou, S. and Axani, S. and Bai, X. and Balagopal V., A. and Barwick, S. W. and Basu, V. and Baur, S. and Bay, R. and Beatty, J. J. and Becker, K.-H. and Tjus, J. Becker and Beise, J. and Bellenghi, C. and Benda, S. and BenZvi, S. and Berley, D. and Bernardini, E. and Besson, D. Z. and Binder, G. and Bindig, D. and Blaufuss, E. and Blot, S. and Boddenberg, M. and Bontempo, F. and Book, J. Y. and Borowka, J. and Böser, S. and Botner, O. and Böttcher, J. and Bourbeau, E. and Bradascio, F. and Braun, J. and Brinson, B. and Bron, S. and Brostean-Kaiser, J. and Burley, R. T. and Busse, R. S. and Campana, M. A. and Carnie-Bronca, E. G. and Chen, C. and Chen, Z. and Chirkin, D. and Choi, K. and Clark, B. A. and Clark, K. and Classen, L. and Coleman, A. and Collin, G. H. and Connolly, A. and Conrad, J. M. and Coppin, P. and Correa, P. and Cowen, D. F. and Cross, R. and Dappen, C. and Dave, P. and De Clercq, C. and DeLaunay, J. J. and López, D. Delgado and Dembinski, H. and Deoskar, K. and Desai, A. and Desiati, P. and de Vries, K. D. and de Wasseige, G. and DeYoung, T. and Diaz, A. and Díaz-Vélez, J. C. and Dittmer, M. and Dujmovic, H. and Dunkman, M. and DuVernois, M. A. and Ehrhardt, T. and Eller, P. and Engel, R. and Erpenbeck, H. and Evans, J. and Evenson, P. A. and Fan, K. L. and Fazely, A. R. and Fedynitch, A. and Feigl, N. and Fiedlschuster, S. and Fienberg, A. T. and Finley, C. and Fischer, L. and Fox, D. and Franckowiak, A. and Friedman, E. and Fritz, A. and Fürst, P. and Gaisser, T. K. and Gallagher, J. and Ganster, E. and Garcia, A. and Garrappa, S. and Gerhardt, L. and Ghadimi, A. and Glaser, C. and Glauch, T. and Glüsenkamp, T. and Goehlke, N. and Goldschmidt, A. and Gonzalez, J. G. and Goswami, S. and Grant, D. and Grégoire, T. and Griswold, S. and Günther, C. and Gutjahr, P. and Haack, C. and Hallgren, A. and Halliday, R. and Halve, L. and Halzen, F. and Minh, M. Ha and Hanson, K. and Hardin, J. and Harnisch, A. A. and Haungs, A. and Helbing, K. and Henningsen, F. and Hettinger, E. C. and Hickford, S. and Hignight, J. and Hill, C. and Hill, G. C. and Hoffman, K. D. and Hoshina, K. and Hou, W. and Huang, F. and Huber, M. and Huber, T. and Hultqvist, K. and Hünnefeld, M. and Hussain, R. and Hymon, K. and In, S. and Iovine, N. and Ishihara, A. and Jansson, M. and Japaridze, G. S. and Jeong, M. and Jin, M. and Jones, B. J. P. and Kang, D. and Kang, W. and Kang, X. and Kappes, A. and Kappesser, D. and Kardum, L. and Karg, T. and Karl, M. and Karle, A. and Katz, U. and Kauer, M. and Kellermann, M. and Kelley, J. L. and Kheirandish, A. and Kin, K. and Kiryluk, J. and Klein, S. R. and Kochocki, A. and Koirala, R. and Kolanoski, H. and Kontrimas, T. and Köpke, L. and Kopper, C. and Kopper, S. and Koskinen, D. J. and Koundal, P. and Kovacevich, M. and Kowalski, M. and Kozynets, T. and Krupczak, E. and Kun, E. and Kurahashi, N. and Lad, N. and Gualda, C. Lagunas and Lanfranchi, J. L. and Larson, M. J. and Lauber, F. and Lazar, J. P. and Lee, J. W. and Leonard, K. and Leszczyńska, A. and Li, Y. and Lincetto, M. and Liu, Q. R. and Liubarska, M. and Lohfink, E. and Mariscal, C. J. Lozano and Lu, L. and Lucarelli, F. and Ludwig, A. and Luszczak, W. and Lyu, Y. and Ma, W. Y. and Madsen, J. and Mahn, K. B. M. and Makino, Y. and Mancina, S. and Mariş, I. C. and Martinez-Soler, I. and Maruyama, R. and McCarthy, S. and McElroy, T. and McNally, F. and Mead, J. V. and Meagher, K. and Mechbal, S. and Medina, A. and Meier, M. and Meighen-Berger, S. and Merckx, Y. and Micallef, J. and Mockler, D. and Montaruli, T. and Moore, R. W. and Morik, K. and Morse, R. and Moulai, M. and Mukherjee, T. and Naab, R. and Nagai, R. and Nahnhauer, R. and Naumann, U. and Necker, J. and Nguyen, L. V. and Niederhausen, H. and Nisa, M. U. and Nowicki, S. C. and Nygren, D. and Pollmann, A. Obertacke and Oehler, M. and Oeyen, B. and Olivas, A. and O’Sullivan, E. and Pandya, H. and Pankova, D. V. and Park, N. and Parker, G. K. and Paudel, E. N. and Paul, L. and de los Heros, C. Pérez and Peters, L. and Peterson, J. and Philippen, S. and Pieper, S. and Pizzuto, A. and Plum, M. and Popovych, Y. and Porcelli, A. and Rodriguez, M. Prado and Pries, B. and Przybylski, G. T. and Raab, C. and Rack-Helleis, J. and Raissi, A. and Rameez, M. and Rawlins, K. and Rea, I. C. and Rechav, Z. and Rehman, A. and Reichherzer, P. and Reimann, R. and Renzi, G. and Resconi, E. and Reusch, S. and Rhode, W. and Richman, M. and Riedel, B. and Roberts, E. J. and Robertson, S. and Roellinghoff, G. and Rongen, M. and Rott, C. and Ruhe, T. and Ryckbosch, D. and Cantu, D. Rysewyk and Safa, I. and Saffer, J. and Salazar-Gallegos, D. and Sampathkumar, P. and Herrera, S. E. Sanchez and Sandrock, A. and Santander, M. and Sarkar, S. and Sarkar, S. and Satalecka, K. and Schaufel, M. and Schieler, H. and Schindler, S. and Schmidt, T. and Schneider, A. and Schneider, J. and Schröder, F. G. and Schumacher, L. and Schwefer, G. and Sclafani, S. and Seckel, D. and Seunarine, S. and Sharma, A. and Shefali, S. and Shimizu, N. and Silva, M. and Skrzypek, B. and Smithers, B. and Snihur, R. and Soedingrekso, J. and Sogaard, A. and Soldin, D. and Spannfellner, C. and Spiczak, G. M. and Spiering, C. and Stamatikos, M. and Stanev, T. and Stein, R. and Stettner, J. and Stezelberger, T. and Stokstad, B. and Stürwald, T. and Stuttard, T. and Sullivan, G. W. and Taboada, I. and Ter-Antonyan, S. and Thwaites, J. and Tilav, S. and Tischbein, F. and Tollefson, K. and Tönnis, C. and Toscano, S. and Tosi, D. and Trettin, A. and Tselengidou, M. and Tung, C. F. and Turcati, A. and Turcotte, R. and Turley, C. F. and Twagirayezu, J. P. and Ty, B. and Elorrieta, M. A. Unland and Valtonen-Mattila, N. and Vandenbroucke, J. and van Eijndhoven, N. and Vannerom, D. and van Santen, J. and Veitch-Michaelis, J. and Verpoest, S. and Walck, C. and Wang, W. and Watson, T. B. and Weaver, C. and Weigel, P. and Weindl, A. and Weiss, M. J. and Weldert, J. and Wendt, C. and Werthebach, J. and Weyrauch, M. and Whitehorn, N. and Wiebusch, C. H. and Willey, N. and Williams, D. R. and Wolf, M. and Wrede, G. and Wulff, J. and Xu, X. W. and Yanez, J. P. and Yildizci, E. and Yoshida, S. and Yu, S. and Yuan, T. and Zhang, Z. and Zhelnin, P.},
   year={2023},
   month=jun, pages={1338–1343} }

@article{ngc1068,
   title={Evidence for neutrino emission from the nearby active galaxy NGC 1068},
   volume={378},
   ISSN={1095-9203},
   url={http://dx.doi.org/10.1126/science.abg3395},
   DOI={10.1126/science.abg3395},
   number={6619},
   journal={Science},
   publisher={American Association for the Advancement of Science (AAAS)},
   author={Abbasi, R. and Ackermann, M. and Adams, J. and Aguilar, J. A. and Ahlers, M. and Ahrens, M. and Alameddine, J. M. and Alispach, C. and Alves, A. A. and Amin, N. M. and Andeen, K. and Anderson, T. and Anton, G. and Argüelles, C. and Ashida, Y. and Axani, S. and Bai, X. and Balagopal V., A. and Barbano, A. and Barwick, S. W. and Bastian, B. and Basu, V. and Baur, S. and Bay, R. and Beatty, J. J. and Becker, K.-H. and Becker Tjus, J. and Bellenghi, C. and BenZvi, S. and Berley, D. and Bernardini, E. and Besson, D. Z. and Binder, G. and Bindig, D. and Blaufuss, E. and Blot, S. and Boddenberg, M. and Bontempo, F. and Borowka, J. and Böser, S. and Botner, O. and Böttcher, J. and Bourbeau, E. and Bradascio, F. and Braun, J. and Brinson, B. and Bron, S. and Brostean-Kaiser, J. and Browne, S. and Burgman, A. and Burley, R. T. and Busse, R. S. and Campana, M. A. and Carnie-Bronca, E. G. and Chen, C. and Chen, Z. and Chirkin, D. and Choi, K. and Clark, B. A. and Clark, K. and Classen, L. and Coleman, A. and Collin, G. H. and Conrad, J. M. and Coppin, P. and Correa, P. and Cowen, D. F. and Cross, R. and Dappen, C. and Dave, P. and De Clercq, C. and DeLaunay, J. J. and Delgado López, D. and Dembinski, H. and Deoskar, K. and Desai, A. and Desiati, P. and de Vries, K. D. and de Wasseige, G. and de With, M. and DeYoung, T. and Diaz, A. and Díaz-Vélez, J. C. and Dittmer, M. and Dujmovic, H. and Dunkman, M. and DuVernois, M. A. and Dvorak, E. and Ehrhardt, T. and Eller, P. and Engel, R. and Erpenbeck, H. and Evans, J. and Evenson, P. A. and Fan, K. L. and Fazely, A. R. and Fedynitch, A. and Feigl, N. and Fiedlschuster, S. and Fienberg, A. T. and Filimonov, K. and Finley, C. and Fischer, L. and Fox, D. and Franckowiak, A. and Friedman, E. and Fritz, A. and Fürst, P. and Gaisser, T. K. and Gallagher, J. and Ganster, E. and Garcia, A. and Garrappa, S. and Gerhardt, L. and Ghadimi, A. and Glaser, C. and Glauch, T. and Glüsenkamp, T. and Goldschmidt, A. and Gonzalez, J. G. and Goswami, S. and Grant, D. and Grégoire, T. and Griswold, S. and Günther, C. and Gutjahr, P. and Haack, C. and Hallgren, A. and Halliday, R. and Halve, L. and Halzen, F. and Ha Minh, M. and Hanson, K. and Hardin, J. and Harnisch, A. A. and Haungs, A. and Hebecker, D. and Helbing, K. and Henningsen, F. and Hettinger, E. C. and Hickford, S. and Hignight, J. and Hill, C. and Hill, G. C. and Hoffman, K. D. and Hoffmann, R. and Hokanson-Fasig, B. and Hoshina, K. and Huang, F. and Huber, M. and Huber, T. and Hultqvist, K. and Hünnefeld, M. and Hussain, R. and Hymon, K. and In, S. and Iovine, N. and Ishihara, A. and Jansson, M. and Japaridze, G. S. and Jeong, M. and Jin, M. and Jones, B. J. P. and Kang, D. and Kang, W. and Kang, X. and Kappes, A. and Kappesser, D. and Kardum, L. and Karg, T. and Karl, M. and Karle, A. and Katz, U. and Kauer, M. and Kellermann, M. and Kelley, J. L. and Kheirandish, A. and Kin, K. and Kintscher, T. and Kiryluk, J. and Klein, S. R. and Koirala, R. and Kolanoski, H. and Kontrimas, T. and Köpke, L. and Kopper, C. and Kopper, S. and Koskinen, D. J. and Koundal, P. and Kovacevich, M. and Kowalski, M. and Kozynets, T. and Kun, E. and Kurahashi, N. and Lad, N. and Lagunas Gualda, C. and Lanfranchi, J. L. and Larson, M. J. and Lauber, F. and Lazar, J. P. and Lee, J. W. and Leonard, K. and Leszczyńska, A. and Li, Y. and Lincetto, M. and Liu, Q. R. and Liubarska, M. and Lohfink, E. and Lozano Mariscal, C. J. and Lu, L. and Lucarelli, F. and Ludwig, A. and Luszczak, W. and Lyu, Y. and Ma, W. Y. and Madsen, J. and Mahn, K. B. M. and Makino, Y. and Mancina, S. and Mariş, I. C. and Martinez-Soler, I. and Maruyama, R. and Mase, K. and McElroy, T. and McNally, F. and Mead, J. V. and Meagher, K. and Mechbal, S. and Medina, A. and Meier, M. and Meighen-Berger, S. and Micallef, J. and Mockler, D. and Montaruli, T. and Moore, R. W. and Morse, R. and Moulai, M. and Naab, R. and Nagai, R. and Nahnhauer, R. and Naumann, U. and Necker, J. and Nguyen, L. V. and Niederhausen, H. and Nisa, M. U. and Nowicki, S. C. and Nygren, D. and Obertacke Pollmann, A. and Oehler, M. and Oeyen, B. and Olivas, A. and O’Sullivan, E. and Pandya, H. and Pankova, D. V. and Park, N. and Parker, G. K. and Paudel, E. N. and Paul, L. and Pérez de los Heros, C. and Peters, L. and Peterson, J. and Philippen, S. and Pieper, S. and Pittermann, M. and Pizzuto, A. and Plum, M. and Popovych, Y. and Porcelli, A. and Prado Rodriguez, M. and Price, P. B. and Pries, B. and Przybylski, G. T. and Raab, C. and Rack-Helleis, J. and Raissi, A. and Rameez, M. and Rawlins, K. and Rea, I. C. and Rehman, A. and Reichherzer, P. and Reimann, R. and Renzi, G. and Resconi, E. and Reusch, S. and Rhode, W. and Richman, M. and Riedel, B. and Roberts, E. J. and Robertson, S. and Roellinghoff, G. and Rongen, M. and Rott, C. and Ruhe, T. and Ryckbosch, D. and Rysewyk Cantu, D. and Safa, I. and Saffer, J. and Sanchez Herrera, S. E. and Sandrock, A. and Sandroos, J. and Santander, M. and Sarkar, S. and Sarkar, S. and Satalecka, K. and Schaufel, M. and Schieler, H. and Schindler, S. and Schmidt, T. and Schneider, A. and Schneider, J. and Schröder, F. G. and Schumacher, L. and Schwefer, G. and Sclafani, S. and Seckel, D. and Seunarine, S. and Sharma, A. and Shefali, S. and Silva, M. and Skrzypek, B. and Smithers, B. and Snihur, R. and Soedingrekso, J. and Soldin, D. and Spannfellner, C. and Spiczak, G. M. and Spiering, C. and Stachurska, J. and Stamatikos, M. and Stanev, T. and Stein, R. and Stettner, J. and Steuer, A. and Stezelberger, T. and Stokstad, R. and Stürwald, T. and Stuttard, T. and Sullivan, G. W. and Taboada, I. and Ter-Antonyan, S. and Tilav, S. and Tischbein, F. and Tollefson, K. and Tönnis, C. and Toscano, S. and Tosi, D. and Trettin, A. and Tselengidou, M. and Tung, C. F. and Turcati, A. and Turcotte, R. and Turley, C. F. and Twagirayezu, J. P. and Ty, B. and Unland Elorrieta, M. A. and Valtonen-Mattila, N. and Vandenbroucke, J. and van Eijndhoven, N. and Vannerom, D. and van Santen, J. and Verpoest, S. and Walck, C. and Watson, T. B. and Weaver, C. and Weigel, P. and Weindl, A. and Weiss, M. J. and Weldert, J. and Wendt, C. and Werthebach, J. and Weyrauch, M. and Whitehorn, N. and Wiebusch, C. H. and Williams, D. R. and Wolf, M. and Woschnagg, K. and Wrede, G. and Wulff, J. and Xu, X. W. and Yanez, J. P. and Yoshida, S. and Yu, S. and Yuan, T. and Zhang, Z. and Zhelnin, P.},
   year={2022},
   month=nov, pages={538–543} }

@article{Neyman:1937uhy,
    author = "Neyman, J.",
    title = "{Outline of a Theory of Statistical Estimation Based on the Classical Theory of Probability}",
    doi = "10.1098/rsta.1937.0005",
    journal = "Phil. Trans. Roy. Soc. Lond. A",
    volume = "236",
    number = "767",
    pages = "333--380",
    year = "1937"
}

@ARTICLE{2019arXiv190105486C,
       author = {{Countryman}, Stefan and {Keivani}, Azadeh and {Bartos}, Imre and {Marka}, Zsuzsa and {Kintscher}, Thomas and {Corley}, Rainer and {Blaufuss}, Erik and {Finley}, Chad and {Marka}, Szabolcs},
        title = "{Low-Latency Algorithm for Multi-messenger Astrophysics (LLAMA) with Gravitational-Wave and High-Energy Neutrino Candidates}",
      journal = {arXiv e-prints},
         year = 2019,
        month = jan,
          eid = {arXiv:1901.05486},
        pages = {arXiv:1901.05486},
          doi = {10.48550/arXiv.1901.05486},
archivePrefix = {arXiv},
       eprint = {1901.05486},
 primaryClass = {astro-ph.HE},
       adsurl = {https://ui.adsabs.harvard.edu/abs/2019arXiv190105486C}
}

@ARTICLE{2013RvMP...85.1401A,
       author = {{Ando}, Shin'ichiro and {Baret}, Bruny and {Bartos}, Imre and {Bouhou}, Boutayeb and {Chassande-Mottin}, Eric and {Corsi}, Alessandra and {Di Palma}, Irene and {Dietz}, Alexander and {Donzaud}, Corinne and {Eichler}, David and {Finley}, Chad and {Guetta}, Dafne and {Halzen}, Francis and {Jones}, Gareth and {Kandhasamy}, Shivaraj and {Kotake}, Kei and {Kouchner}, Antoine and {Mandic}, Vuk and {M{\'a}rka}, Szabolcs and {M{\'a}rka}, Zsuzsa and {Moscoso}, Luciano and {Papa}, Maria Alessandra and {Piran}, Tsvi and {Pradier}, Thierry and {Romero}, Gustavo E. and {Sutton}, Patrick and {Thrane}, Eric and {Van Elewyck}, V{\'e}ronique and {Waxman}, Eli},
        title = "{Colloquium: Multimessenger astronomy with gravitational waves and high-energy neutrinos}",
      journal = {Reviews of Modern Physics},
         year = 2013,
        month = oct,
       volume = {85},
       number = {4},
        pages = {1401-1420},
          doi = {10.1103/RevModPhys.85.1401},
archivePrefix = {arXiv},
       eprint = {1203.5192},
 primaryClass = {astro-ph.HE},
       adsurl = {https://ui.adsabs.harvard.edu/abs/2013RvMP...85.1401A}
}

@ARTICLE{2012PhRvD..85j3004B,
       author = {{Baret}, Bruny and {Bartos}, Imre and {Bouhou}, Boutayeb and {Chassande-Mottin}, Eric and {Corsi}, Alessandra and {Di Palma}, Irene and {Donzaud}, Corinne and {Drago}, Marco and {Finley}, Chad and {Jones}, Gareth and {Klimenko}, Sergey and {Kouchner}, Antoine and {M{\'a}rka}, Szabolcs and {M{\'a}rka}, Zsuzsa and {Moscoso}, Luciano and {Papa}, Maria Alessandra and {Pradier}, Thierry and {Prodi}, Giovanni and {Raffai}, Peter and {Re}, Virginia and {Rollins}, Jameson and {Salemi}, Francesco and {Sutton}, Patrick and {Tse}, Maggie and {Van Elewyck}, V{\'e}ronique and {Vedovato}, Gabriele},
        title = "{Multimessenger science reach and analysis method for common sources of gravitational waves and high-energy neutrinos}",
      journal = {Physical Review D},
         year = 2012,
        month = may,
       volume = {85},
       number = {10},
          eid = {103004},
        pages = {103004},
          doi = {10.1103/PhysRevD.85.103004},
archivePrefix = {arXiv},
       eprint = {1112.1140},
 primaryClass = {astro-ph.HE},
       adsurl = {https://ui.adsabs.harvard.edu/abs/2012PhRvD..85j3004B}
}

@ARTICLE{2009IJMPD..18.1655V,
       author = {{van Elewyck}, Veronique and {Ando}, S. and {Aso}, Y. and {Baret}, B. and {Barsuglia}, M. and {Bartos}, I. and {Chassande-Mottin}, E. and {di Palma}, I. and {Dwyer}, J. and {Finley}, C. and {Kei}, K. and {Kouchner}, A. and {Marka}, S. and {Marka}, Z. and {Rollins}, J. and {Ott}, C.~D. and {Pradier}, T. and {Searle}, A.},
        title = "{Joint Searches Between Gravitational-Wave Interferometers and High-Energy Neutrino Telescopes:. Science Reach and Analysis Strategies}",
      journal = {International Journal of Modern Physics D},
         year = 2009,
        month = jan,
       volume = {18},
       number = {10},
        pages = {1655-1659},
          doi = {10.1142/S0218271809015655},
archivePrefix = {arXiv},
       eprint = {0906.4957},
 primaryClass = {astro-ph.IM},
       adsurl = {https://ui.adsabs.harvard.edu/abs/2009IJMPD..18.1655V}
}

@ARTICLE{2008CQGra..25k4039A,
       author = {{Aso}, Yoichi and {M{\'a}rka}, Zsuzsa and {Finley}, Chad and {Dwyer}, John and {Kotake}, Kei and {M{\'a}rka}, Szabolcs},
        title = "{Search method for coincident events from LIGO and IceCube detectors}",
      journal = {Classical and Quantum Gravity},
         year = 2008,
        month = jun,
       volume = {25},
       number = {11},
          eid = {114039},
        pages = {114039},
          doi = {10.1088/0264-9381/25/11/114039},
archivePrefix = {arXiv},
       eprint = {0711.0107},
 primaryClass = {astro-ph},
       adsurl = {https://ui.adsabs.harvard.edu/abs/2008CQGra..25k4039A}
}

@article{PhysRevLett.107.251101,
  volume = {107},
  journal = {Physical Review Letters},
  month = {Dec},
  numpages = {4},
  author = {Bartos, Imre and Finley, Chad and Corsi, Alessandra and M\'arka, Szabolcs},
  title = {Observational Constraints on Multimessenger Sources of Gravitational Waves and High-Energy Neutrinos},
  year = {2011},
  url = {http://link.aps.org/doi/10.1103/PhysRevLett.107.251101},
  doi = {10.1103/PhysRevLett.107.251101},
  issue = {25},
  publisher = {American Physical Society},
  pages = {251101}
}

@ARTICLE{2013JCAP...06..008A,
   author = {{Adri{\'a}n-Mart{\'{\i}}nez}, S. and {Samarai}, I.~A. and {Albert}, A. and
	{Andr{\'e}}, M. and {Anghinolfi}, M. and {Anton}, G. and {Anvar}, S. and
	{Ardid}, M. and {Astraatmadja}, T. and {Aubert}, J.-J. and et al.},
    title = "{A first search for coincident gravitational waves and high energy neutrinos using LIGO, Virgo and ANTARES data from 2007}",
  journal = {\jcap},
archivePrefix = "arXiv",
   eprint = {1205.3018},
 primaryClass = "astro-ph.HE",
     year = 2013,
    month = jun,
   volume = 6,
      eid = {008},
    pages = {8},
      doi = {10.1088/1475-7516/2013/06/008},
   adsurl = {http://adsabs.harvard.edu/abs/2013JCAP...06..008A}
}

@article{Abbott_2016,
   title={Observation of Gravitational Waves from a Binary Black Hole Merger},
   volume={116},
   ISSN={1079-7114},
   url={http://dx.doi.org/10.1103/PhysRevLett.116.061102},
   DOI={10.1103/physrevlett.116.061102},
   number={6},
   journal={Physical Review Letters},
   publisher={American Physical Society (APS)},
   author={Abbott, B. P. and Abbott, R. and Abbott, T. D. and Abernathy, M. R. and Acernese, F. and Ackley, K. and Adams, C. and Adams, T. and Addesso, P. and Adhikari, R. X. and others},
   year={2016},
   month=feb }

@ARTICLE{2011APh....35....1B,
   author = {{Baret}, B. and {Bartos}, I. and {Bouhou}, B. and {Corsi}, A. and
	{Palma}, I.~D. and {Donzaud}, C. and {Elewyck}, V.~V. and {Finley}, C. and
	{Jones}, G. and {Kouchner}, A. and {M{\'a}rka}, S. and {M{\'a}rka}, Z. and
	{Moscoso}, L. and {Chassande-Mottin}, E. and {Papa}, M.~A. and
	{Pradier}, T. and {Raffai}, P. and {Rollins}, J. and {Sutton}, P.
	},
    title = "{Bounding the time delay between high-energy neutrinos and gravitational-wave transients from gamma-ray bursts}",
  journal = {Astroparticle Physics},
archivePrefix = "arXiv",
   eprint = {1101.4669},
 primaryClass = "astro-ph.HE",
     year = 2011,
    month = aug,
   volume = 35,
    pages = {1-7},
      doi = {10.1016/j.astropartphys.2011.04.001},
   adsurl = {http://adsabs.harvard.edu/abs/2011APh....35....1B}
}

@INPROCEEDINGS{2019ICRC...36.1021B,
       author = {{Blaufuss}, E. and {Kintscher}, T. and {Lu}, L. and {Tung}, C.~F.},
        title = "{The Next Generation of IceCube Real-time Neutrino Alerts}",
    booktitle = {36th International Cosmic Ray Conference (ICRC2019)},
         year = 2019,
       series = {International Cosmic Ray Conference},
       volume = {36},
        month = jul,
          eid = {1021},
        pages = {1021},
          doi = {10.22323/1.358.1021},
archivePrefix = {arXiv},
       eprint = {1908.04884},
 primaryClass = {astro-ph.HE},
       adsurl = {https://ui.adsabs.harvard.edu/abs/2019ICRC...36.1021B}
}

@article{PhysRevD.47.2198,
  title = {Observing binary inspiral in gravitational radiation: One interferometer},
  author = {Finn, Lee Samuel and Chernoff, David F.},
  journal = {Physical Review D},
  volume = {47},
  issue = {6},
  pages = {2198--2219},
  numpages = {0},
  year = {1993},
  month = {Mar},
  publisher = {American Physical Society},
  doi = {10.1103/PhysRevD.47.2198},
  url = {https://link.aps.org/doi/10.1103/PhysRevD.47.2198}
}

@article{Chen_2021,
doi = {10.1088/1361-6382/abd594},
url = {https://dx.doi.org/10.1088/1361-6382/abd594},
year = {2021},
month = {jan},
publisher = {IOP Publishing},
volume = {38},
number = {5},
pages = {055010},
author = {Hsin-Yu Chen and Daniel E Holz and John Miller and Matthew Evans and Salvatore Vitale and Jolien Creighton},
title = {Distance measures in gravitational-wave astrophysics and cosmology},
journal = {Classical and Quantum Gravity}
}

@ARTICLE{2021ApJ...909..116A,
       author = {{Abe}, S. and {Asami}, S. and {Gando}, A. and {Gando}, Y. and {Gima}, T. and {Goto}, A. and {Hachiya}, T. and {Hata}, K. and {Hayashida}, S. and {Hosokawa}, K. and {Ichimura}, K. and {Ieki}, S. and {Ikeda}, H. and {Inoue}, K. and {Ishidoshiro}, K. and {Kamei}, Y. and {Kawada}, N. and {Kishimoto}, Y. and {Kinoshita}, T. and {Koga}, M. and {Maemura}, N. and {Mitsui}, T. and {Miyake}, H. and {Nakamura}, K. and {Nakamura}, K. and {Nakamura}, R. and {Ozaki}, H. and {Sakai}, T. and {Sambonsugi}, H. and {Shimizu}, I. and {Shirai}, J. and {Shiraishi}, K. and {Suzuki}, A. and {Suzuki}, Y. and {Takeuchi}, A. and {Tamae}, K. and {Ueshima}, K. and {Wada}, Y. and {Watanabe}, H. and {Yoshida}, Y. and {Obara}, S. and {Kozlov}, A. and {Chernyak}, D. and {Takemoto}, Y. and {Yoshida}, S. and {Umehara}, S. and {Fushimi}, K. and {Ichikawa}, A.~K. and {Nakamura}, K.~Z. and {Yoshida}, M. and {Berger}, B.~E. and {Fujikawa}, B.~K. and {Learned}, J.~G. and {Maricic}, J. and {Axani}, S.~N. and {Winslow}, L.~A. and {Fu}, Z. and {Ouellet}, J. and {Efremenko}, Y. and {Karwowski}, H.~J. and {Markoff}, D.~M. and {Tornow}, W. and {Li}, A. and {Detwiler}, J.~A. and {Enomoto}, S. and {Decowski}, M.~P. and {Grant}, C. and {O'Donnell}, T. and {Dell'Oro}, S. and {KamLAND Collaboration}},
        title = "{Search for Low-energy Electron Antineutrinos in KamLAND Associated with Gravitational Wave Events}",
      journal = {The Astrophysical Journal},
         year = 2021,
        month = mar,
       volume = {909},
       number = {2},
          eid = {116},
        pages = {116},
          doi = {10.3847/1538-4357/abd5bc},
archivePrefix = {arXiv},
       eprint = {2012.12053},
 primaryClass = {astro-ph.HE},
       adsurl = {https://ui.adsabs.harvard.edu/abs/2021ApJ...909..116A}
}
\bibliographystyle{aasjournal}
\appendix

\begin{table}[h]
    \centering
    \begin{tabular}{|r|r|r|r|r|r|r|}
    \hline
    GW event GPS Time & $p$-value [$10^{-3}$] & Energy [GeV] & R.A. [deg] & Dec. [deg] & 90\% containment radius [deg] & $\Delta t$ [s]\\ \hline
    \textbf{CBC Events} &\,&\,&\,&\,&\,& \\ \hline
         1262142545.615 & 0.38 & 2416 & 37.31 & 7.46  & 2.62 & $-$222\\ \hline
         1258769976.671 & 0.52 & 2553 & 213.01 & 24.91 & 2.40 & $-$232\\ \hline
         1260567236.512 & 0.52 & 890  & 323.19 & 4.53  & 4.08 & $-$43\\ \hline
         1264693411.565 & 0.83 & 1045 & 138.52 & 14.12 & 1.59 & $-$59\\ \hline
         1261317209.144 & 1.1 & 4260 & 11.03  & 53.31 & 5.23 & 412\\ \hline
         1240986884.531 & 1.3 & 1222 & 161.03 & 39.41 & 3.41 & 3\\ \hline
         1248922975.872 & 1.7 & 1099 & 11.80  & 43.66 & 0.86 & $-$79\\ \hline
         1251996593.946 & 1.9 & 6656 & 341.15 & 3.53  & 0.43 & 85\\ \hline
         1261395060.713 & 2.0 & 1227 & 7.32   & 25.41 & 2.17 & 55\\ \hline
         1242369559.803 & 2.3 & 4938 & 67.79  & 80.85 & 6.24 & $-$148\\ \hline
         1245646385.592 & 2.6 & 2615 & 181.93 & $-$5.55 & 1.07 & $-$131\\ \hline
         1259556411.479 & 3.5 & 921  & 342.74 & 51.25 & 2.98 & $-$309\\ \hline
         1257878102.660 & 3.8 & 1549 & 78.01  & 15.62 & 2.17 & $-$286\\ \hline
         1262800092.537 & 5.6 & 702  & 289.72 & 0.23  & 1.22 & 85\\ \hline
         1257317013.751 & 5.7 & 1264 & 91.41  & 35.29 & 3.07 & $-$209\\ \hline
         1248959867.938 & 5.8 & 2090 & 62.89  & 29.51 & 2.60 & 25\\ \hline
         1239281492.824 & 6.1 & 1385 & 60.72  & 19.89 & 1.03 & 49\\ \hline
         1246655423.219 & 7.9 & 16865 & 277.26 & $-$7.61 & 0.43 & 295\\ \hline
         1258085887.274 & 8.3 & 1708 & 23.33  & 17.38 & 2.29 & 3\\ \hline
         1238533707.234 & 8.3 & 2344 & 299.04 & 22.97 & 1.37 & 23\\ \hline
         1249072561.974 & 9.3 & 1986 & 125.07 & 31.54 & 4.65 & 296\\ \hline
         1262751997.732 & 9.6 & 1448 & 81.07  & 55.20 & 1.27 & $-$46\\ \hline
        \textbf{cWB Events}  &\,&\,&\,&\,&\,& \\ \hline
        1241247887.938&0.43&2268&242.35&51.34&3.20 &154\\ \hline
        1259085330.234&1.1&1960&197.90&25.78&3.13&6\\ \hline
         1266181040.000&1.3&2636&144.50&7.87&1.27&389\\ \hline
         1258325594.125&4.3&330&7.44&19.58&4.03&171\\ \hline
         1250185147.797&8.9&710&131.95&11.31&2.04&-19\\ \hline

    \end{tabular}
    \caption{Properties of the coincident neutrinos that produce a $p$-value of 0.01 or less. Sky coordinates are in right ascension (R.A.) and declination (Dec.). The time difference $\Delta t$ is neutrino detection time minus GW event time.}
    \label{tab:nutable}
\end{table}

\end{document}